\documentclass[aps,prb,showpacs,amssymb,amsfonts,amsmath,groupedaddress,twocolumn]{revtex4-1}
\usepackage{bm}
\usepackage{amsmath}
\usepackage{wrapfig}
\usepackage[dvips]{graphicx}
\usepackage{longtable}
\usepackage{overpic}
\begin{document}

%\preprint{}

\title{{\large Numerical Investigation of Triexciton Stabilization in Diamond with Multiple Valleys and Bands}}


\author{Hiroki Katow$^1$, Junko Usukura$^2$, Ryosuke Akashi$^1$, K\'alm\'an Varga$^3$, Shinji Tsuneyuki$^1$}
%\thanks{}
\affiliation{{\small $^1$The University of Tokyo, 7-3-1 Hongo, Bunkyo-ku, Tokyo 113-8656, Japan}}
\affiliation{{\small $^2$Department of Physics, Tokyo University of Science, 1-3 Kagurazaka, Tokyo 162-8601, Japan}}
\affiliation{{\small $^3$Department of Physics and Astronomy, Vanderbilt University, Nashville, Tennessee 37235, United States}}

\date{\today}

\begin{abstract}
 The existence of polyexcitons, the $N$-body complexes of excitons for $N > 2$ in 3D bulk systems, has been controversial for more than 40 years since its first theoretical suggestion. We investigated the stability of fundamental excitonic complexes in diamond numerically with the stochastic variational method (SVM) and an explicitly correlated Gaussian (ECG) basis. The electron-hole many-body system is described by an effective mass Hamiltonian.  Our model includes the effective mass anisotropy and multiple valley and band degrees of freedom. We show that the excitons, trions, biexcitons, charged biexcitons, and triexcitons are stable in diamond. Numerical calculations reproduce from 81\% to 86\% of the experimentally reported binding energies for neutral bound states. 
\end{abstract}


\maketitle

\section{Introduction}
\label{sec:intro}
A wide variety of electron-hole bound states appears in photo-excited semiconductors due to the attractive or repulsive Coulomb interaction between the carriers. The fundamental composite particles are the excitons (e$^-$ + hole), charged excitons (trion, exciton + e$^-$(hole)), and biexcitons. The polyexcitons (PE$_n$), excitonic $n$-body complexes are considered to be one of such various electron-hole many-body bound states. Although there have been some theoretical studies calculating the binding energies of charged biexcitons and smaller complexes in bulk or 2D systems\cite{CG_excitonic}$^\text{-}$\cite{BiexTMD2}, numerical investigations of PE$_n$ ($n > 2$) are still missing except in quantum dots\cite{PEinQDot}. In this paper, we report the first numerical evidence for triexciton stability resulting from the effects of multiple valley and band degrees of freedom and large effective mass anisotropy of diamond.

The identification of excitonic complexes has a great significance since they play an essential role in the optical response of solids. Large excitonic effects in the photoluminescence or photoabsorption spectrum can be seen not only in bulk systems\cite{Wolfe}$^,$\cite{Nagai} but also in 2D systems like MoS$_2$\cite{RevTMD}$^,$\cite{TrionMoS2}$^,$\cite{IntValleyBiex}, WSe$_2$\cite{BiexWSe2}, and WS$_2$\cite{BiexWS2}, or lower dimensional systems like quantum dots\cite{QDot}. Complexes such as the exciton and biexciton  form insulating gas phases in electron-hole many-body systems. Phase diagrams of such systems are roughly estimated for fundamental semiconductors like silicon \cite{Wolfe} or diamond \cite{Nagai}. Identification of possible excitonic bound states is essential to establish the phase diagrams. 

%polyexciton
It is not easy to show the stability of PE$_n$ for $n > 2$, or in general, complex particles of strongly interacting negatively and positively charged particles. For example, the positronium trimer (Ps$_3$) was shown to be unstable by a precise numerical calculation\cite{TriPs}, and the hydrogen trimer (H$_3$) is also known to be unstable\cite{TriH}. This fact suggests that PE$_n$ for $n > 2$ are unstable in direct gap semiconductors. However, in the case of indirect gap semiconductors, it has been predicted that degenerate valleys and valence bands relax the Pauli repulsion between identical particles and make PE$_n$ stable. The possible existence of PE$_n$ was first pointed out by Wang and Kittel\cite{Kittel}, and they estimated the binding energy of polyexcitons in the heavy hole limit $m_e \ll m_h$, where $m_e$ $(m_h)$ is the effective mass of the electron (hole). 

%excitonic comlexes
The concept of the PE$_n$ in bulk systems has been well accepted, and experimental signatures of polyexcitons are reported in the silicon and diamond. Steele, McMullan, and Thewalt first observed a series of peaks in the four-particle decay process (two electron-hole pairs decay into one photon) in high-purity silicon\cite{SiPoly}, and they attributed the peaks to biexciton decays in PE$_n$ up to $n = 4$. However, the indistinct line shape of the spectra and the small binding energy of the exciton brought about controversial discussions including the alternative interpretation of the observed spectra by a new kind of electron-hole plasma\cite{Wolfe}$^,$\cite{SiPoly}$^\text{-}$\cite{SCP2}. Recently in diamond, Omachi {\it et al.}\cite{polyexciton} reported six photoemission peaks energetically lower than the single exciton peak and they attributed these peaks to exciton decay in PE$_n (n = 2 - 6)$. Here, the peak positions were precisely observed thanks to the large exciton binding energy in diamond. 

%excitonic comlexes
 Numerical simulations of electron-hole systems with more than six particles are challenging, and there are few theoretical studies of polyexcitons. One reason is the high computational cost of solving strongly correlated few-body problems with more than 6 particles, and other reasons are the complexities of degenerate multiple valley and band and the large effective mass anisotropy. Cancio and Chang reported numerical calculations up to PE$_4$ by the Quantum Monte Carlo method\cite{QMC}. They employed a spherical effective mass model and a trial function that was symmetrized under the permutation of electrons(holes). This symmetrization cannot be justified since it makes the wave function unchanged under a permutation of identical particles. It possibly overestimates the binding energy since the exchange interaction between identical particles becomes attractive. 
 
%Description of few-body system
 The Hylleraas-type basis function or James-Coolidge-type functions\cite{James-Coolidge} are frequently used to express the wave function of quantum few-body systems. 
One reason is that the value of $\{\Psi^{-1} (\partial \Psi/\partial r)\}_{r = 0}$, where $\Psi$ is the wave function and $r$ is an inter-particle distance, is equal to the known exact value. This is known as the cusp condition.
%they satisfy the cusp condition which provides the exact value of $\{\Psi^{-1} (\partial \Psi/\partial r)\}_{r = 0}$, where $\Psi$ is the wave function and $r$ is an inter-particle distance. 
 Another is that they reproduce the long range behavior of the wave function of Coulombic few body system, decaying as $e^{-\alpha r}$, where $\alpha$ is constant. On the other hand, they require numerical integrations to calculate matrix elements of the Coulomb potential. The other possible candidate is the explicitly-correlated-Gaussian (ECG) type basis. The ECG type basis function is provided by a Gaussian which depends on relative coordinates like
\begin{eqnarray}
\psi(\bm{r}) \propto \exp \{-\sum_{i < j}^N\frac{1}{2}A_{ij}(\bm{r}_i-\bm{r}_j)^2\},
\label{eq:ECG}
\end{eqnarray}
where $A_{ij}$ is a variational parameter which determines the width of the Gaussian. A great advantage of the ECG type function is that all matrix elements of Hamiltonian of the Coulombic system can be calculated analytically. Considering the fact that the analytical form of the cusp condition is unknown in strongly anisotropic systems like diamond, the ECG basis possibly becomes a more powerful choice.

%outline
In this paper, we report the stability of excitonic complexes in diamond up to the triexcitons. Our results are the first numerical evidence for the triexciton stability in bulk systems in the sense that the bound states are calculated by diagonalizing directly the few-body Hamiltonian with anisotropic effective masses and degenerate valley and band degrees of freedom. The effective mass Hamiltonian, the ECG basis function for the trial wave function, and a stochastic method for parameter optimization are introduced in Sec.~II. We report the ground state binding energies and separation energies of the excitons, trions (charged exciton), biexcitons, charged biexcitons(CBE), and triexcitons in Sec.~III. The separation energy is defined as an energy to separate an exciton from a PE$_n$, trion, and CBE. Our conclusions are summarized in Sec.~IV.
%\subsection{BACKGROUND}
%\subsubsection{}
%\subsection{Purpose}
\section{THEORETICAL FORMULATION}

\subsection{Model Hamiltonian}
\label{sec:Hamiltonian}
 In the Brillouin zone of diamond, the valence band maximum is triply degenerate at the $\Gamma$ point belonging to $\Gamma_{25'}$ representation of the $O_h$ group. The conduction band has six energetically equivalent minima (valleys) on the $\Delta$ axes which belong to the $\Delta_1$ representation of the $C_{4v}$ group (See Fig.~\ref{fig:Decay}(b)). Hereafter we use the notation $\Gamma_{xy}$, $\Gamma_{yz}$, $\Gamma_{zx}$ for the triply degenerate valence bands and $\Delta_{(\pm k_c 0 0)}$, $\Delta_{(0 \pm k_c 0)}$, $\Delta_{(0 0 \pm k_c)}$ for six equivalent valleys, where the subscripts are the coordinates of the valleys in the Brillouin zone. We employed the following $\bm{k}\cdot \bm{p}$ effective mass Hamiltonian for a general $N$-body electron-hole system in diamond with multiple valley and band. 
\begin{eqnarray}
\mathcal{H} = \sum_{i = 1}^{N_h}\sum_{\Gamma}t_{h,i}^{(\Gamma)}\cdot |\Gamma_i\rangle \langle \Gamma_i|
	+ \sum_{i}^{N_h}\sum_{\Gamma,\Gamma' } t_{h,i}^{(\Gamma \Gamma')}\cdot |\Gamma_i\rangle \langle \Gamma'_i|\nonumber \\
	+\sum_{i = 1}^{N_e}\sum_{\Delta}t_{e,i}^{(\Delta)} \cdot |\Delta_i\rangle \langle \Delta_i|	+\sum_{i,j}^{N_e+N_h}V_{ij}
	\label{eq:Hamiltonian}
\end{eqnarray}
The first and third terms are the kinetic energy of the holes and electrons, respectively. The second term is the inter-band coupling of the holes. The last term is the isotropic Coulomb interaction which is screened by the dielectric constant $\epsilon$ as follows :
\begin{equation}
 V_{ij} =  \frac{e_ie_j}{\epsilon r_{ij}}.
\end{equation}
$e_{i(j)}$ is the charge of $i(j)$-th particle and $r_{ij} = |\bm{r}_i-\bm{r}_j|$. In the present notation, the subscripts $\Gamma$ and $\Delta$ run over valley and band degrees of freedom, respectively. $|\gamma_i\rangle$ ($\gamma = \Delta,\; \Gamma$) is the Bloch function of $i$ th particle which satisfies the orthonormality relation :
\begin{eqnarray}
 \langle \gamma_i | \gamma'_ i \rangle = \delta_{\gamma_i \gamma'_i}.
	\label{eq:tau}
\end{eqnarray}
Here we show an example of how to evaluate the kinetic energy term $\sum_{\gamma}t_{h,i}^{(\gamma)}\cdot |\gamma_i\rangle \langle \gamma_i|$ for the trial wave function $|\psi\rangle = f(\bm{r}) \cdot \prod_{i = 1}^{N_e}|\Delta_i\rangle\prod_{i = 1}^{N_h}|\Gamma_i\rangle $ :
\begin{eqnarray}
&&\langle \psi '|\sum_{\gamma}t_{h,i}^{(\gamma)}\cdot |\gamma_i\rangle \langle \gamma_i||\psi\rangle \nonumber\\
&&=  \sum_{\gamma}\int d\bm{r}^{3N}f'(\bm{r})t_{h,i}^{(\gamma)}f(\bm{r})  \nonumber\\
	&&\times \prod_{j = 1}^{N_e}\langle\Delta'_j|\prod_{j = 1}^{N_h}\langle\Gamma'_j | |\gamma_i\rangle \langle \gamma_i|\prod_{j = 1}^{N_e}|\Delta_j\rangle\prod_{j = 1}^{N_h}|\Gamma_j\rangle \nonumber\\
&&=\int d\bm{r}^{3N}f'(\bm{r})t_{h,i}^{(\Gamma_i)}f(\bm{r}) \cdot \delta_{\Gamma'_i\Gamma_i} \nonumber\\
	&&\times \prod_{j = 1}^{N_e}\delta_{\Delta'_j\Delta_j}\prod_{j (\neq i)}^{N_h}\delta_{\Gamma'_j\Gamma_j}.
\end{eqnarray}
Here, $f$ is the envelope function factor and $|\Delta\rangle$ and $|\Gamma\rangle$ are the Bloch function factors of the trial wave function. We excluded the spin function for simplicity. 
The analytical form of the kinetic energy terms is given by
\begin{eqnarray}
	\left\{ \begin{array}{ll}
    	t_{h,i}^{(\Gamma_{yz})} = L\partial_x^2+M(\partial_y^2 + \partial_z^2) \\
	t_{h,i}^{(\Gamma_{zx})} = L\partial_y^2+M(\partial_x^2 + \partial_z^2) \\
	t_{h,i}^{(\Gamma_{xy})} = L\partial_z^2+M(\partial_x^2 + \partial_y^2)\\	
	\end{array} \right.
	\label{eq:kinetic_h}	
\end{eqnarray}
\begin{eqnarray}
	&&\left\{ \begin{array}{ll}
    	t_{e,i}^{(\Delta_{(\pm k_c00)})} = -\frac{1}{2}\{ \frac{1}{m_l}(\partial_x\pm ik_{c})^2+\frac{1}{m_t}(\partial_y^2 + \partial_z^2) \}\\
	t_{e,i}^{(\Delta_{(0\pm k_c 0)})} = -\frac{1}{2}\{ \frac{1}{m_l}(\partial_y\pm ik_{c})^2+\frac{1}{m_t}(\partial_x^2 + \partial_z^2) \} \\
	t_{e,i}^{(\Delta_{(00\pm k_c)})} = -\frac{1}{2}\{ \frac{1}{m_l}(\partial_z\pm ik_{c})^2+\frac{1}{m_t}(\partial_x^2 + \partial_y^2) \}.	
	\end{array} \right.
	\label{eq:kinetic_e}
\end{eqnarray}
Similarly, the inter-band coupling terms are given by
\begin{eqnarray}
	\left\{ \begin{array}{ll}
    	t_{h,i}^{(\Gamma_{yz}\Gamma_{zx})} = N\partial_x\partial_y\\
	t_{h,i}^{(\Gamma_{zx}\Gamma_{xy})} = N\partial_y\partial_z\\
	t_{h,i}^{(\Gamma_{xy}\Gamma_{yz})} = N\partial_z\partial_x,	
	\end{array} \right.
	 \label{eq:inter-band}
\end{eqnarray} 
where $\partial_x$, $\partial_y$, and $\partial_z$ are partial derivatives with respect to the $x$, $y$ and $z$ components of the one-particle coordinates.  We used experimental values of the effective mass parameters and the dielectric constant \cite{mass}. $L = -2.06$, $M = -4.48$, $N = 5.32$ in units of $\frac{\hbar^2}{2m_0}$, where $m_0$ is the free electron mass.  Similarly, $m_l = 1.56 m_0, m_t = 0.280m_0$, and the dielectric constant is $\epsilon = 5.70$. In Eq.~(\ref{eq:kinetic_h}) and (\ref{eq:kinetic_e}), the effective mass along one axis is heavier than along the other directions. Each valley or band is distinguished by the direction of this axis.   
  
 The spin-orbit splitting of top of the valence band, the inter-valley scattering effect, and the electron-hole exchange interaction are neglected in our model for simplicity. In diamond, the spin-orbit splitting is $6$ meV and the electron-hole exchange interaction is also of the same order\cite{exchange}, and these values are relatively small compared with the observed binding energy of exciton in diamond (80 meV\cite{exciton}). This smallness justifies our approximations which neglect these two effects.  

\subsection{Trial Wave Function}
\label{sec:WaveFunction}
To overcome the complexities due to the strongly anisotropic environment of diamond and the high computational cost of calculating the PE$_n$ eigenfunction by frequently used methods, we use the explicitly correlated Gaussian (ECG) basis in Eq.~(\ref{eq:ECG}) for the trial wave function. As we mentioned above, the Hylleraas-type function or James-Coolidge-type functions\cite{James-Coolidge} are frequently used for the description of quantum Coulombic  few-body systems. These basis states explicitly depend on the inter-particle distances {\it e.g.} $r_{12} = |\bm{r}_1-\bm{r}_2|$ like
\begin{eqnarray}
\psi_k \propto r_{12}\exp \{-\alpha_k r_{12}\}.
\end{eqnarray}
Their characteristic properties are that (i)they satisfy the exact value of the derivative of wave function at the origin of inter-particle distance $\{(\partial \Psi/\partial r)/\Psi\}_{r = 0}$, {\it i.e.} the cusp condition, and (ii)they reproduce the exponential decay of the wave function  of the Coulombic few-body system at large inter-particle distances. In particular (i) is expected to greatly reduce the number of basis states.  A significant disadvantage of using this type of basis state is the high computational cost due to the numerical integrations needed for calculations of matrix elements. The analytical form of the exact cusp condition is unknown for anisotropic systems. 
Therefore, it is uncertain that the property (i) reduces the number of basis states in our case.
On the other hand, the ECG basis state reproduces the value of the exact wave function in the vicinity of origin and in the long range limit with a sufficient number of basis states\cite{CGtext}, although the ECG basis state in Eq.~(\ref{eq:ECG}) does not satisfy the exact cusp condition.
This property is expected to be valid even in the case of the anisotropic Hamiltonian in Eq.(\ref{eq:Hamiltonian}).
The ECG basis enables us to calculate all matrix elements of the anisotropic Hamiltonian Eq.~(\ref{eq:Hamiltonian}) analytically, and therefore the ECG basis should be superior for time consuming calculations of PE$_n$ $(n > 2)$ in strongly anisotropic systems.

For the description of excitonic $N$-body bound states, we express the trial wave function $|\Psi\rangle$ in terms of a non-orthogonal basis set as $|\Psi\rangle = \sum_{k} C_k |\psi_k\rangle$. $C_k$ is a real expansion coefficient. Each $|\psi_k\rangle$ is factorized into the envelope function $f_k(\bm{r})$, the spin function  $\chi_{sm_s}$, and Bloch functions $|\Delta_i\rangle$ and $|\Gamma_i\rangle$ as follows :
\begin{eqnarray}
|\psi_k\rangle = \mathcal{A}\{f_k(\bm{r})\cdot \chi_{sm_s} \cdot \prod_{i = 1}^{N_e}|\Delta_i\rangle\prod_{i = 1}^{N_h}|\Gamma_i\rangle \}.
\label{eq:basis}
\end{eqnarray}
The Bloch function of the $i$ th electron(hole), $|\Delta_i\rangle$ ($|\Gamma_i\rangle$), satisfies the orthonormality relation in Eq.~(\ref{eq:tau}).  $\mathcal{A}$ is an antisymmetrizer operating on the trial wave function so that the Pauli exclusion principle is satisfied. To construct the envelope function, we introduce  the Jacobi coordinate $\mathrm{x}_i$ as a set of relative coordinates which is defined by
\begin{eqnarray}
&&\mathrm{x}_i = \sum_{j = 1}^{N_e + N_h} U_{ij}\bm{r}_j \\
&&U = \begin{pmatrix}
1 & -1 & 0 & \cdot \cdot \cdot &0 \\
m_1/M_2 &m_2/M_2& -1 &\cdot \cdot \cdot&0  \\
m_1/M_3 &m_2/M_3& \cdot \cdot \cdot &\cdot \cdot \cdot&0  \\
\cdot &\cdot&&&\cdot  \\
\cdot &\cdot&&&\cdot  \\
\cdot &\cdot&&&\cdot  \\
m_1/M_N &m_2/M_N& \cdot \cdot \cdot &\cdot \cdot \cdot&m_N/M_N  \\
\end{pmatrix},
\end{eqnarray}
where $M_n$ is a sum of the geometric mean of the effective mass, {\it i.e.} $M_n = m_1 + m_2 + \cdot\cdot\cdot +m_n$ and $m_i = \{m_{i,x}m_{i,y}m_{i,z}\}^{1/3}$ where $m_{i,j}$ ($j = x,y,z$) is the effective mass of $i$ th particle along direction $j$ and $\bm{r}_i$ is an one-particle coordinate. The envelope function is constructed from the ECG basis\cite{CGauss} by using the Jacobi coordinates as follows :
\begin{eqnarray}
	f_k(\bm{r}) = \theta_{L}(\bm{v_k})\cdot \exp\{-\frac{1}{2}\mathrm{xA_kx}\} \prod_{i}^{N_e}\exp\{i\bm{k_{\Delta_i}\cdot r_i}\} .
	\label{eq:envelope}
\end{eqnarray}
Here $\theta_{L}(\bm{v_k})$ is the non-spherical factor and consists of a real solid spherical harmonic that depends on the {\it global vector} $\bm{v} = \sum_{i = 1}^{N_e+N_h-1}u_{k,i}\mathrm{x}_i$ \cite{gvector}. The coefficient $u_{k,i}$ is a variational parameter. 
\begin{eqnarray}
   \theta_{L}(\bm{v}) = \left\{ \begin{array}{ll}
    	|\bm{v}|^L \{Y_{LM}(\hat{\bm{v}})+Y_{L-M}(\hat{\bm{v}}) \} & (M > 0) \\
    	i|\bm{v}|^L \{Y_{LM}(\hat{\bm{v}})-Y_{L-M}(\hat{\bm{v}}) \} & ( M < 0) \\
    	|\bm{v}|^L Y_{LM}(\hat{\bm{v}})& (M = 0)
  \end{array} \right.
  \label{eq:theta}
\end{eqnarray}
In Eq.~(\ref{eq:theta}), $Y_{LM}(\hat{\bm{v}})$ is a spherical harmonic depending on the direction of $\bm{v}$, {\it i.e.} $\hat{\bm{v}} = \bm{v}/|\bm{v}|$.  In the envelope function Eq.~ (\ref{eq:envelope}), the Gaussian factor is characterized by the variational parameter $\mathrm{A}_{k,ij}$ where $\mathrm{xA_kx} = \sum_{i,j}^{N_e+N_h-1} \mathrm{x}_i\mathrm{A}_{k,ij}\mathrm{x}_j$.  $\bm{k}_{\Delta_i}$ is the constant wave vector of valley $\Delta_i$.  Our trial function does not depend on the center-of-mass coordinate $\mathrm{x}_N$ $(N = N_e+N_h)$. The ECG basis has been used to obtain accurate energies for many different few-body systems\cite{CGauss}. Superposition of different angular-momentum states is essential to express the orbital deformation of the envelope function caused by the effective mass anisotropy.  Due to the plane wave factor, the contribution from the valley wave vector completely vanishes in the matrix elements of the Hamiltonian in Eq.(\ref{eq:Hamiltonian}). We fixed the total spin of the trial wave function as a singlet in neutral (or even particle number) systems and $1/2$ in charged (or odd particle number) systems. 

%In our calculations, $k_c$ in (\ref{eq:kinetic_e}), the absolute value of valley wave vector, is completely canceled out by the plane wave part of the trial wave function in Eq.(\ref{eq:envelope}). Therefore, the matrix elements of the Hamiltonian do not depend on $k_c$ in our calculations (See Appendix).

\subsection{Parameter optimization}
\label{sec:optimization}
To optimize  the trial wave function, we employed the stochastic variational method (SVM)\cite{SVM}. In the SVM, the number of basis states is increased one by one up to an arbitrary number, and in each step the variational parameters $u_{k,i}$ and  $\mathrm{A}_{k,ij}$ are determined by random sampling. The original SVM consists of the following two steps. 

(i){\it Increasing process.} Assume that there are $K$ basis states with the ground state energy $E_K$. In the increasing process, first, $P$ random basis states are generated. New ground state energies are calculated by using existing $K$ basis states and one of generated $P$ basis states. Then new ground state energies $E_i$ ($i=1, \cdot \cdot \cdot , P$) are given. The $(K+1)$ th basis state is chosen so that it gives the lowest ground state energy, and is added to the existing $K$ basis states. This increasing process is repeated until the number of basis states reaches a certain number which is chosen so that the binding energy converges within the desired accuracy. In our case, we repeatedly increased the maximum number of basis states until predetermined accuracy was achieved. 

(ii){\it Refinement process.} In the refinement process, the $K$-dimensional basis set, that is determined in the increasing process, is improved by replacing the basis states by better ones with the total number of basis states fixed. $P'$ basis states are randomly generated as new candidates for the $k$-th basis states and new ground state energies $E_i$ ($i=1, \cdot \cdot \cdot , P'$) are calculated, in which the $k$-th basis state is replaced by the candidates. If the energy which is lowest in the newly calculated $E_i$ ($i=1, \cdot \cdot \cdot , P'$) is lower than the original states, then the existing $k$-th basis state is replaced with the new basis state. This procedure is repeated for $k = 1, \cdot \cdot \cdot , K$.

By using step (ii), an energy improvement less than 0.1\% is reported for a ground state calculation of positronium molecule\cite{CGdist}. We obtained the convergence of binding energies within 1 \% in the following results without imposing the step (ii), and it is sufficiently accurate to show the stability of excitonic complexes.  Typical numbers of basis states in our calculations range from 50 to 100 in excitons and from 1000 to 1800 in triexcitons.  

%\subsection{Intrinsic Density Distribution}
%\subsection{Correlation Function}
\section{RESULTS and DISCUSSIONS}
\subsection{Contribution of the inter-band coupling to exciton binding energies}
\label{sec:interband}
To examine the accuracy of the present method, we first applied it to excitons in GaN with anisotropic effective masses in one valley and one band. We defined the exciton binding energy as the energy to separate an exciton into a free electron-hole pair. We obtained 24.809 meV for a heavy-hole exciton, 15.445 meV for a light-hole exciton, respectively. These values are in excellent agreement with earlier theoretical binding energies obtained by exact diagonalization of a single-band electron-hole effective mass Hamiltonian\cite{GaN}: 24.809 meV for the heavy-hole exciton, 15.458 meV for the light-hole exciton, respectively. 

\begin{figure}[htbp]
 % \begin{center}
    \begin{tabular}{cc}
      % 1
      \begin{minipage}{1.0\hsize}
        %\begin{center}
          %\includegraphics[clip, width=7cm,bb=0 0 891 1191]{ExTri2.jpg}%{exciton.eps}
          %\includegraphics[clip, width=6cm]{Dissociate.eps}%{exciton.eps}
          \begin{overpic}[clip, width=7cm]{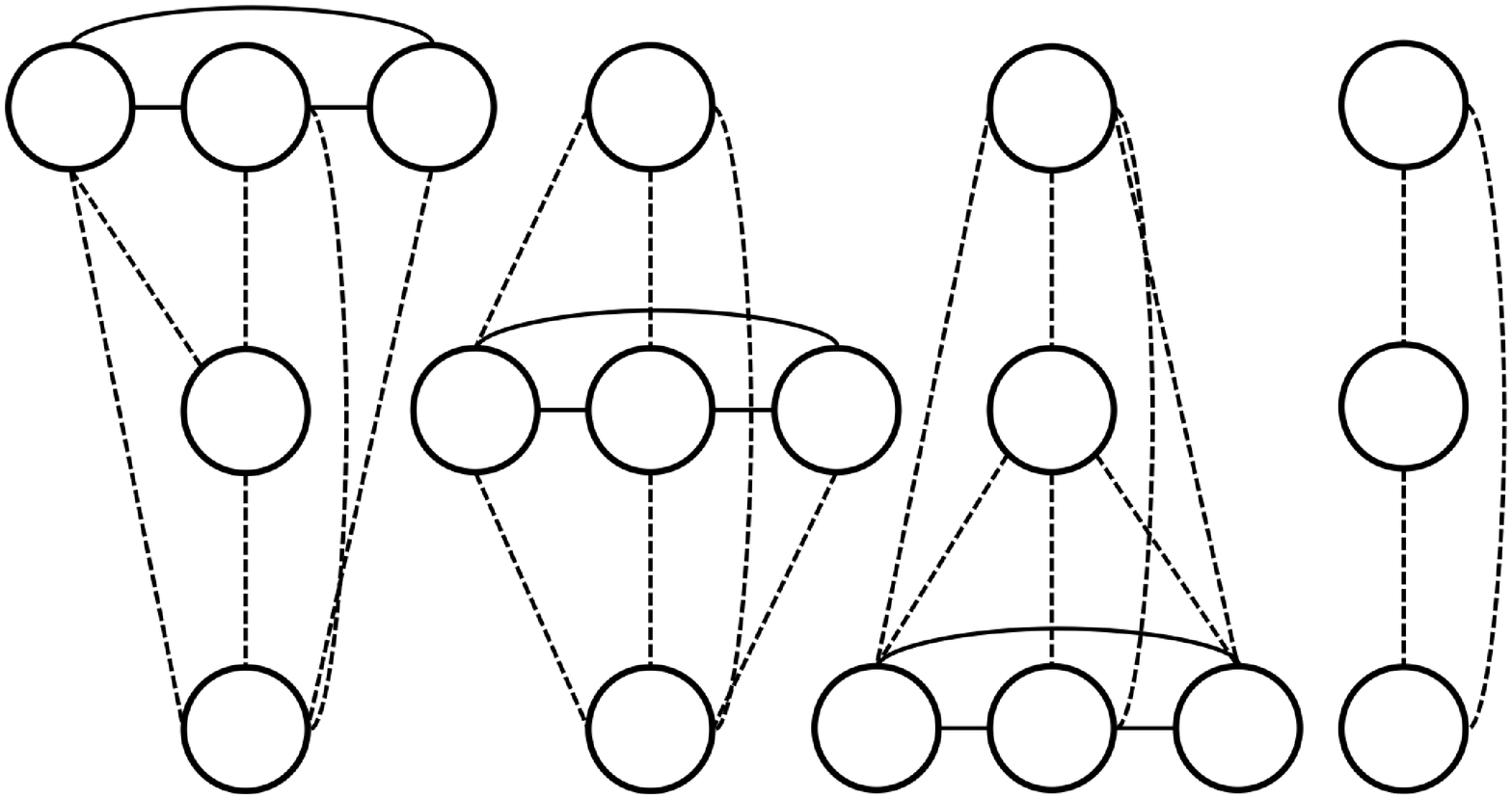}%{exciton.eps}
          %\begin{overpic}[clip, width=7cm]{coupling.pdf}
	  \put(-15,52){(a)}
	  \put(-15,45){$|\Gamma_{yz}\rangle$}	  
	  \put(3,45){$s$}	  
	  \put(9.5,39.5){$d_{z^2+c}$}	  
	  \put(21.5,39.5){$d_{x^2-y^2}$}	  
	  \put(40,45){$d_{xy}$}	  
	  \put(67,45){$d_{zx}$}	  
	  \put(90,45){$d_{yz}$}	  
	  \put(-15,25){$|\Gamma_{zx}\rangle$}	  
	  \put(13,25){$d_{xy}$}	  
	  \put(30,25){$s$}	  
	  \put(37,19){$d_{z^2+c}$}	  
	  \put(50,19){$d_{x^2-y^2}$}	  
	  \put(67,25){$d_{yz}$}	  
	  \put(90,25){$d_{zx}$}	  
	  \put(-15,3){$|\Gamma_{xy}\rangle$}	  
	  \put(13,3){$d_{zx}$}	  
	  \put(40,3){$d_{yz}$}	  
	  \put(57,3){$s$}	  
	  \put(63,-2){$d_{z^2+c}$}	  
	  \put(75.5,-2){$d_{x^2-y^2}$}	  
	  \put(90,3){$d_{xy}$}	  
	  \end{overpic}
          \hspace{1.6cm}
        %\end{center}
      \end{minipage}\\

      % 2
      \begin{minipage}{0.6\hsize}
        \begin{center}
         \begin{overpic}[clip, width=5cm]{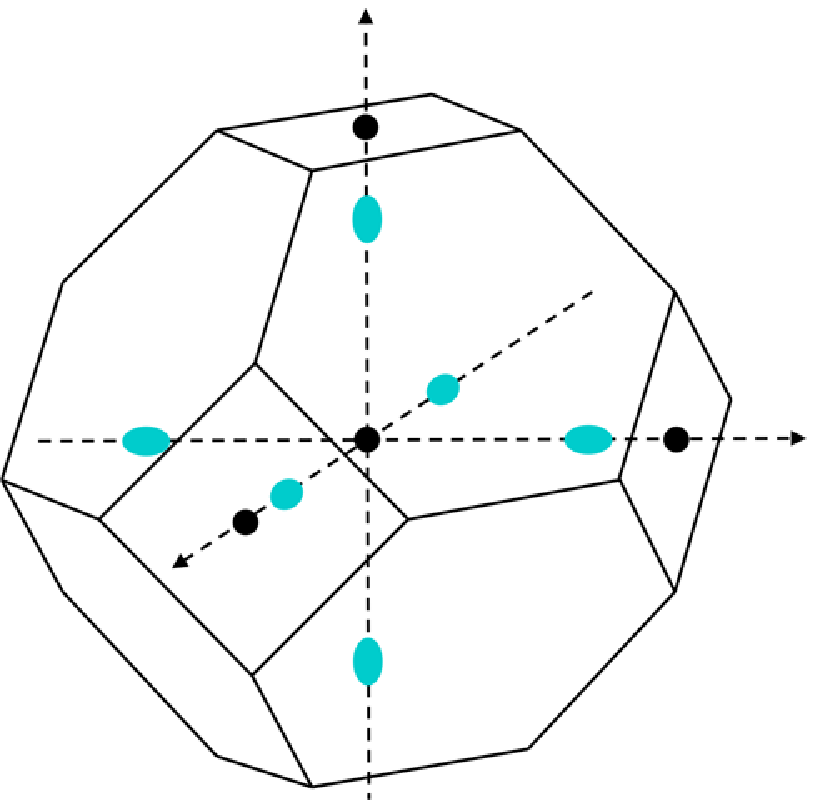}%{exciton.eps}
          %\begin{overpic}[clip, width=5cm]{BZ.pdf}
	  \put(0,90){(b)}
  	  \put(37.5,48){$\Gamma_{25'}$}
    	  \put(37,90){$k_z$}
    	  \put(22,35){$k_x$}	 
    	  \put(90,48){$k_y$}	 
	  \put(68,48){$\Delta_{1}$}
	  \end{overpic}
          %\hspace{1.6cm}
        \end{center}
      \end{minipage}

      % 3
      \begin{minipage}{0.4\hsize}
        \begin{center}
          \begin{overpic}[clip, width=5cm, angle = -90]{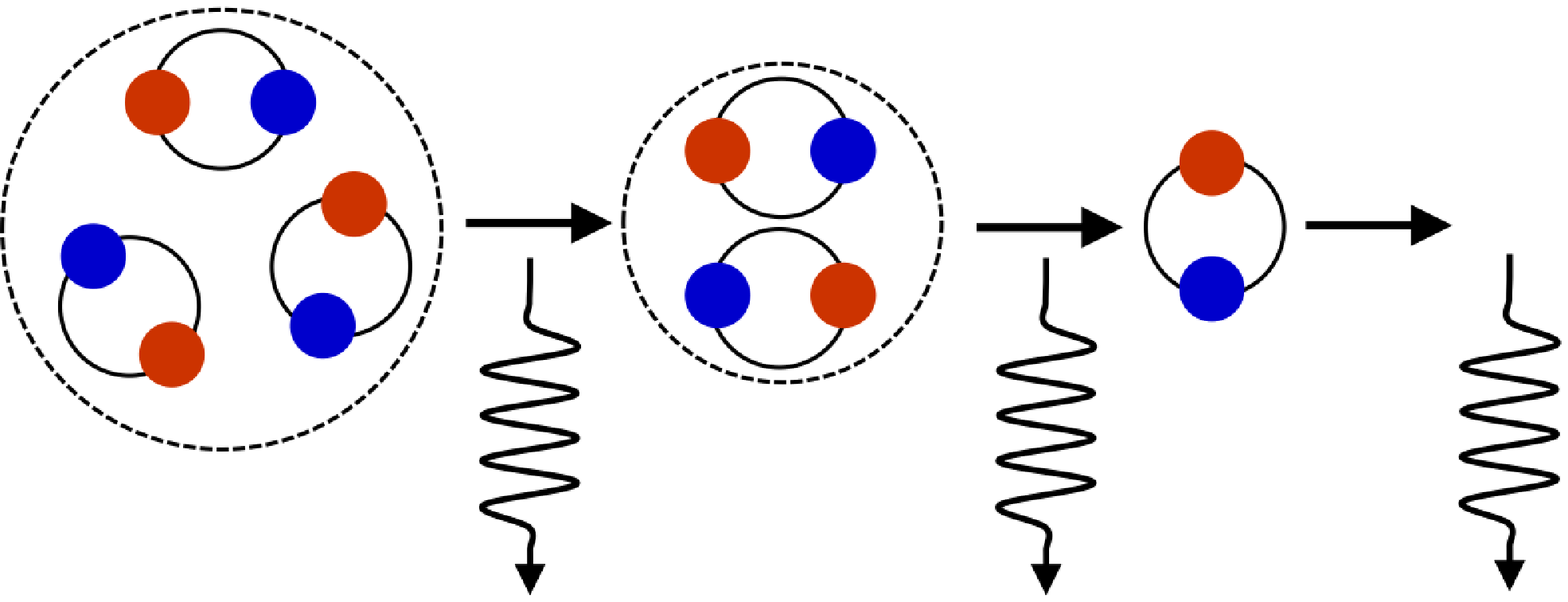}%{exciton.eps}
          %\begin{overpic}[clip, width=5cm, angle = -90]{Dissociate.pdf}%{exciton.eps}
	  \put(-10,90){(c)}
	  \put(42,85){PE$_3$}
	  \put(40,47){PE$_2$}
	  \put(34,21.5){exciton}
	  \put(0,70){X$_3$}
	  \put(0,37){X$_2$}
	  \put(0,7){EX}
	  \end{overpic}
          %\hspace{1.6cm}
        \end{center}
      \end{minipage}
	
    \end{tabular}
    \caption{(Color online) (a)Diagram showing couplings between orbitals by the kinetic energy (solid lines) and inter-band coupling (dotted lines) of the Hamiltonian Eq.~(\ref{eq:Hamiltonian}). Orbitals in the first, second and third rows belong to the $|\Gamma_{yz}\rangle$, $|\Gamma_{zx}\rangle$ and $|\Gamma_{xy}\rangle$ bands.  (b)The first Brillouin zone of diamond. The six blue spots are valleys on $\Delta$ axes. (c)Schematic picture of exciton decay in polyexcitons(PE$_n$). X$_3$ and X$_2$ are photons emitted from exciton recombination in PE$_3$ and PE$_2$, respectively. EX is photoemission from a free exciton recombination. }
    \label{fig:Decay}
  %\end{center}
\end{figure}

Before showing the results for excitons in diamond, we discuss the contribution of the inter-band coupling Eq.~(\ref{eq:inter-band}) to the binding energy. It is clear from symmetry that $s$-type orbitals are not coupled with one another by the inter-band coupling. From parity conservation, $d$-type orbitals are the lowest angular-momentum states which can couple with $s$-type orbitals. Matrix elements of the inter-band coupling between two angular momentum eigenstates $|lm\rangle$ and $|l'm'\rangle$ are non-zero only if $|l'm'\rangle$ satisfies following relations.
\begin{eqnarray}
   \left\{ \begin{array}{ll}
    	t_{h,i}^{(\Gamma_{yz}\Gamma_{zx})}  &: |l'm'\rangle = |l\;m\pm2\rangle, \; |l\pm2\;m\pm2\rangle\\
    	t_{h,i}^{(\Gamma_{xz}\Gamma_{xy})}  &: |l'm'\rangle = |l\;m\pm1\rangle, \; |l\pm2\;m\pm1\rangle \\
    	t_{h,i}^{(\Gamma_{xy}\Gamma_{yz})}  &: |l'm'\rangle = |l\;m\pm1\rangle, \; |l\pm2\;m\pm1\rangle 
  \end{array} \right.
\end{eqnarray}
Fig.~\ref{fig:Decay}(a) shows $s$- and five $d$-type orbitals coupled with one another by the kinetic energy term (solid lines) and inter-band coupling terms (dotted lines) of the Hamiltonian Eq.(\ref{eq:Hamiltonian}). Matrix elements of the Hamiltonian Eq.~(\ref{eq:Hamiltonian}) between two orbitals are non-zero only if the two orbitals are connected by a solid or dotted line in Fig.~\ref{fig:Decay}(a). We compare the following two trial functions to see the contribution of the inter-band coupling terms.
\begin{eqnarray}
   \left\{ \begin{array}{ll}
	(\mathrm{i})  \{f_{s}+f_{d_{z^2+c}}+f_{d_{x^2-y^2}}\}|\Delta_{(0k_c0)}\rangle|\Gamma_{xy}\rangle\\
				\;\;\;+ f_{d_{zx}}|\Delta_{(0k_c0)}\rangle|\Gamma_{yz}\rangle
				+ f_{d_{xy}}|\Delta_{(0k_c0)}\rangle|\Gamma_{zx}\rangle\\
				\\
	(\mathrm{ii})  \{f_{s}+f_{d_{z^2+c}}+f_{d_{x^2-y^2}}\}|\Delta_{(00k_c)}\rangle|\Gamma_{yz}\rangle
   \end{array} \right.
   \label{eq:2type}
\end{eqnarray}
Here $f$ is the envelope function and $|\Delta\rangle$ and $|\Gamma\rangle$ are factors of the Bloch function of trial wave function. The subscripts of the envelope function $f$ denote the symmetry of the real solid spherical harmonic. Eq.~(\ref{eq:2type})(i) corresponds to the second cluster from the right hand side in Fig.~\ref{fig:Decay}(a) and it is fully connected by the kinetic energy and inter-band coupling. Eq.~(\ref{eq:2type})(ii) corresponds to the same cluster but with the inter-band coupling neglected. In the above expressions, we omitted the spin function for simplicity.   We obtained the binding energy from the trial wave function in Eq.~(\ref{eq:2type})(i) as 67.7 meV and in Eq.~(\ref{eq:2type})(ii) as 67.9 meV . The energy improvement from the inter-band coupling is only 0.3 \% and the contributions from the second and third terms in Eq.~(\ref{eq:2type})(i) are negligible. In the following calculations, we therefore neglect the inter-band coupling. Then each bound state is characterized by a single product of valleys and bands like in Eq.~(\ref{eq:2type})(ii). Hereafter we denote a combination of valleys and bands by, for example $\Delta_{(00k_c)}/\Gamma_{yz}$ in the case of Eq.~(\ref{eq:2type})(ii).

\subsection{Experimentally observed binding energy of polyexciton}
\label{sec:experiment}
Now we introduce our method of estimating the binding energies of polyexcitons from the experimentally observed photoluminescence spectrum. Omachi {\it et al.}\cite{polyexciton} observed five peaks ($\mathrm{X}_2$ -- $\mathrm{X}_6$) below the peak of the free exciton recombination(EX) in the photoluminescence spectrum. They attributed X$_n$($n$ = 2 -- 6) to exciton decays in PE$_n$. Fig.~\ref{fig:Decay}(c) is a schematic diagram of the process. The energy gap 
\begin{equation}
S_{\text{PE}_n} = E_{\text{X}_n} - E_\text{EX}
\end{equation}
 is interpreted as the energy to separate PE$_n$ into PE$_{n-1}$ and an isolated exciton. Here $E_{\text{X}_n}$ and $E_{\text{EX}}$ are energies of the peak X$_n$ and a photon emitted from the free exciton recombination process, respectively. According to the paper of Omachi {\it et al.}, $S_{\text{PE}_2}/R = 0.15$, and $S_{\text{PE}_3}/R = 0.31$, where $R = 80$ meV is the binding energy of a free exciton. Then the binding energy of a biexciton $E_{\text{biex}}$ and a triexciton $E_{\text{triex}}$ can be calculated as $E_{\text{biex}} = 2R + S_{\text{PE}_2} = 172$ meV, $E_{\text{triex}} =  E_{\text{biex}} + R + S_{\text{PE}_3} = 277$ meV. 

\subsection{Stability of Excitonic Complexes in diamond}
\label{sec:polyexciton}
\begin{figure}[htbp]
  \begin{center}
    \begin{tabular}{c}

      % 1
      \begin{minipage}{1.0\hsize}
        \begin{center}
          \includegraphics[clip, width=8.5cm]{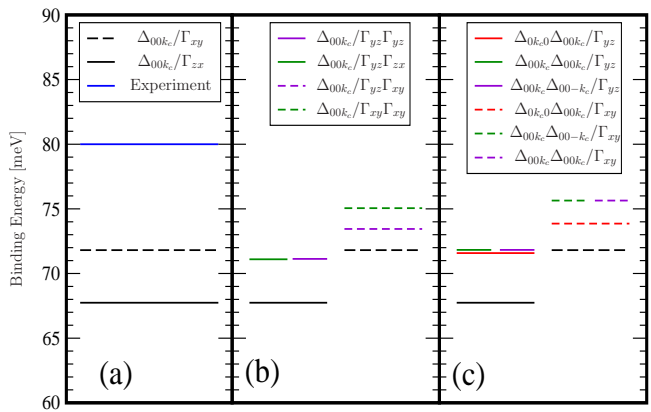}
          \hspace{1.6cm}
        \end{center}
      \end{minipage}\\
	
    \end{tabular}
    \caption{(Color online) (a)Binding energies of excitons. Black lines are calculated values. Blue solid line is the experimentally observed value (Ref.~\onlinecite{exciton}). (b)Binding energies of trion$^+$ (exciton + hole). Black solid lines show threshold energy of dissociation into a pair of a free exciton and a hole for $\Delta_{00k_c}/\Gamma_{yz}\Gamma_{yz}$ and $\Delta_{00k_c}/\Gamma_{yz}\Gamma_{zx}$. Black broken lines show threshold energies of dissociation into a pair of free exciton and a hole for $\Delta_{00k_c}/\Gamma_{yz}\Gamma_{xy}$ and $\Delta_{00k_c}/\Gamma_{xy}\Gamma_{xy}$. Colored lines are calculated values. (c)Binding energies of trion$^-$ (exciton + electron). Black solid lines and black broken lines are threshold energies of dissociation into a pair of free exciton and an electron. Colored lines are calculated values.}
    \label{fig:ExcitonTrion}
  \end{center}
\end{figure}

We show the binding energies of excitons in Fig.~\ref{fig:ExcitonTrion}(a). Here, we defined the binding energy as that required to separate excitonic complexes into free electrons and holes. We obtained 71.8 meV for the combination of $\Delta_{00k_c}/\Gamma_{xy}$ (black broken line) and 67.7 meV for the combination of $\Delta_{00k_c}/\Gamma_{zx}$(black solid line). The energy gap between these two states is attributed to the difference of effective mass anisotropy between the $\Gamma_{xy}$ band and the $\Gamma_{zx}$ band. These results amount to 90\% and 85\% of the experimental value (80 meV\cite{exciton}), respectively. This is enough precision to obtain bound states of triexciton as we will see later. The energy discrepancy between the calculated and experimental values is possibly due to the approximations of our model in which the spin-orbit splitting and electron-hole exchange interaction are excluded. The inter-valley scattering effect may also be present, while the order of contribution in this discrepancy is not clear.

 In the case of the trion, we depicted the threshold energies in Fig.~\ref{fig:ExcitonTrion} to make it easy to see the stability of the trion. If the binding energy of the trion is smaller than the threshold energy, the trion is unstable and spontaneously dissociates into an exciton and a free electron or a free hole.  The binding energies and threshold energies of the trion$^+$ and trion$^-$ are shown in Fig.~\ref{fig:ExcitonTrion}(b) and (c), respectively.
 The number of possible combinations of inequivalent valleys and bands under the spatial symmetry operations are six for trion$^+$ and four for trion$^-$. The average binding energies are 72.6 meV for trion$^+$ and 73.1 meV for trion$^-$. These results show that trion binding energies are not sensitive to the large difference between the electron and hole effective mass. We also depicted the threshold energies for trion$^+$ and trion$^-$ by black solid and broken lines in Fig.~\ref{fig:ExcitonTrion}.  The average binding energy of the trion$^+$ and trion$^-$ are 105\% and 106\% of the exciton binding energy, and these values are close to the results of an earlier theoretical study for trions with isotropic effective mass\cite{CG_excitonic}. There is no experimentally observed value for the trion in diamond, to our knowledge.  

\begin{figure}[htbp]
  \begin{center}
    \begin{tabular}{c}
      % 1
      \begin{minipage}{1.0\hsize}
        \begin{center}
          \includegraphics[clip, width=8.0cm, angle = 0]{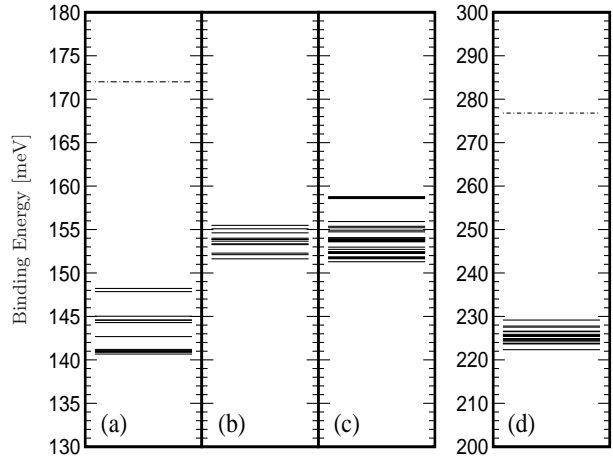}
          \hspace{1.6cm} 
        \end{center}
      \end{minipage}

    \end{tabular}
    \caption{Binding energies of (a)biexciton, (b)CBE$^+$(biexciton + hole), (c) CBE$^-$(biexciton + electron), and (d)triexciton. Solid lines are the calculated binding energy. The dashed-dotted line is the binding energy of the biexciton  calculated from experimentally observed peak positions in the photoluminescence spectrum (Ref.~\onlinecite{polyexciton}).}
    \label{fig:Biexciton}
  \end{center}
\end{figure}

Next, we show the binding energies of biexcitons, charged biexcitons(CBE$^+$ and CBE$^-$), and triexcitons in Fig.~\ref{fig:Biexciton}(a)--(d). Depending on the combination of valleys and bands,  the binding energies of biexcitons vary from 140 meV to 148 meV, and amount to 83\% of the experimentally observed value on average.  In the case of triexcitons, the binding energies reproduce 81\% of the experimental value on average, and vary from 223 meV to 229 meV.  The calculations of biexciton binding energies showed that all combinations of valleys and bands are stable against dissociation into a pair of excitons. The stabilization of charged biexcitons and triexcitons in 3D bulk systems has a great significance. In direct gap semiconductors without any band degeneracy, the triexcitons are predicted to be unstable because of the Pauli blocking effect. The charged biexciton is also unstable unless the condition ($\sigma$ = $m_e/m_h < 0.2$) is satisfied according to an earlier accurate numerical calculation\cite{CG_excitonic}. If we take the geometric mean of effective masses of electron and hole in diamond as $m^* = \{m_xm_ym_z\}^{1/3}$, the effective mass ratio becomes $\sigma = m_h^*/m_e^* =  0.58$ and does not satisfy this condition. Our results of triexciton and charged biexciton binding energies in Fig.~\ref{fig:Biexciton}(d) are the first numerical evidence supporting the existence of the triexcitons and charged biexcitons in a semiconductor with multiple valley and band degrees of freedom. It should be also mentioned that we obtained unbound states in the case of the charged biexciton and triexciton when the three identical particles occupy one valley or band. This calculation condition corresponds to the case of a single band. Therefore, our results do not contradict with those of the earlier calculation\cite{CG_excitonic} and indicate that the stability of charged biexcitons and triexcitons originates from the multiple valley and band degrees of freedom. The next important point is the variation of ground state energy level caused by the effective mass anisotropy of the respective valleys or bands. The width of the energy level distribution varies from about 4 meV to 8 meV depending on the combinations of particles. We expect that the width of the binding energy distribution contributes to  the form of the experimental photoemission spectra. To see this effect, we discuss the separation energies in the next subsection. 

\subsection{Separation energies}
\label{sec:Separation}
\begin{figure}[htbp]
	\begin{center}
 	\includegraphics[clip, width=8.5cm, angle = 0]{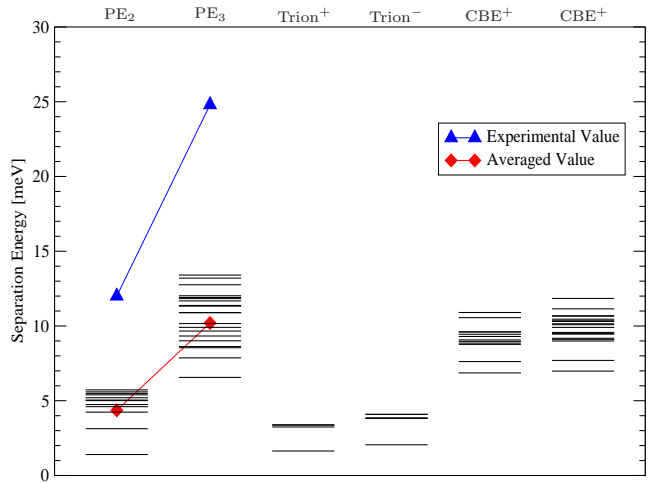}	
	\hspace{1.6cm} 
	\caption{(Color online) Separation energies of PE$_n$(n = 2,3,4), Trions (trion$^{+, -}$)), and Charged Biexcitons (CBE$^{+, -}$)). The black lines are values calculated from the theoretical binding energies. The red diamonds are averaged values. The blue triangles are the experimentally observed values (Ref.~\onlinecite{polyexciton}).}
	\label{fig:Separation}
	\end{center}
\end{figure}

To compare with the experimentally observed photoluminescence spectra, we calculate the separation energies. The separation energy $S_{\text{PE}_n}$ is defined as the minimum energy needed to separate PE$_n$ into a PE$_{n-1}$ and a free exciton as follows:
\begin{eqnarray}
E_{\text{PE}_n} = E_{\text{PE}_{n-1}} + E_{\text{exciton}} + S_{\text{PE}_n} ,
\end{eqnarray}
where $E_{\text{PE}_n}$, $E_{\text{PE}_{n-1}}$, and $E_{\text{exciton}}$ are binding energies of a PE$_n$, PE$_{n-1}$, and exciton, respectively. In the cases of trion and CBE, 
\begin{eqnarray}
E_{\text{trion}^\pm}  = E_{\text{e$^-$(hole)}} + E_{\text{exciton}} + S_{\text{trion}^\pm}\\
E_{\text{CBE}^\pm}  = E_{\text{trion}^\pm} + E_{\text{exciton}}+ S_{\text{CBE}^\pm}.
\end{eqnarray}
Here $E_{\text{e$^-$(hole)}}$ is the energy of free electron or hole, and we set it to zero in our calculation. $E_{\text{trion}^\pm}$ and $E_{\text{CBE}^\pm }$ are the binding energies of the trion$^\pm$ and CBE$^\pm$, respectively. $S_{\text{PE}_n}$ is as defined in Sec.~\ref{sec:experiment} and has a one-to-one correspondence with the energy gaps between experimentally observed peak positions in exciton recombination spectra $X_n$ in PE$_n$. Calculated separation energies are shown in Fig.~\ref{fig:Separation}.
The black solid lines are separation energies of PE$_n$ and charged bound states. We can see widely distributed separation energies in the width of few meV in Fig.~\ref{fig:Separation} as is seen in Fig.~\ref{fig:Biexciton}. Our calculations underestimate the separation energies of PE$_n$ and reproduce 36\% and 41\% of the experimental values for PE$_2$ and PE$_3$, respectively. See Table \ref{tb:separation}. The ratio $S_{\text{PE}_3}/S_{\text{PE}_2}$ gives 113\% of the experimental value and hence our results coincide qualitatively with the experimental values. $S_{\text{PE}_2}$ can be interpreted as the energy of inter-exciton bonding. The fact that $S_{\text{PE}_3}$ is almost two times larger than $S_{\text{PE}_2}$ suggests a simple picture in which $\text{PE}_3$ is a particle in which 3 excitons are weakly bound with each other. 
\begin{table}[htbp]
\begin{tabular}{cccc} \hline
&$S_{\text{PE}_2}\text{[meV]}$&$S_{\text{PE}_3}\text{[meV]}$& $S_{\text{PE}_3}/S_{\text{PE}_2}$ \\ \hline\hline
Calculated Value& $4.36$ & $10.2$&2.33 \\ %\hline
Experimental Value &$12.0$ &$24.8$&2.06  \\ \hline
\end{tabular}
\caption{ Comparison of calculated and experimentally observed value (Ref.~\onlinecite{polyexciton}) of the separation energy $S_{\text{PE}_n}$ for $n = 2, 3$. }
\label{tb:separation}
\end{table}

We can make two predictions by comparing the separation energy distribution with the photoluminescence spectra. 
First, the experimentally observed photoluminescence components of PE$_2$ and PE$_3$ ($i.\;e.$ X$_2$ and X$_3$) may also contain components originating from the trion and CBE, respectively. Secondly, in addition to the temperature of the excitonic complex gas, the separation energy distribution also contributes to the peak width of EX and X$_n$. 
%widthのことについて、なぜこの幅が生まれるかっていうことと、この幅が発光スペクトルのピーク幅の解釈をどう変更するかということをもっと詳しく。
This is because the distribution of the separation energy is roughly interpreted as that of the peak position of the X$_n$ in the photoluminescence spectrum, since the separation energies correspond to the energy gap between the peak positions of the free exciton emission (EX) and the complex particle decay (X$_n$).
In particular the second point will make it difficult to extract the temperature of the excitonic complex gas from the peak width of X$_n$.

\section{CONCLUSION}
In this paper, we investigated the stability of excitonic complexes in diamond by numerical calculation.  The electron-hole system in diamond was described by a $\bm{k}\cdot \bm{p}$ effective mass Hamiltonian with multiple valley and band and Coulomb interaction. The spin-orbit splitting and  electron-hole exchange interaction were neglected for simplicity. The ECG basis, which has been used for precise calculations of binding energies in many few-body systems\cite{CGauss}, was employed for the trial wave function.

Numerical simulations show the stability of triexciton and charged biexciton. The stability of these bound states is one of the most remarkable consequences of the multiple valley and band degrees of freedom in a 3D bulk system. We obtained 81\% to 90\% of the experimentally observed binding energies of PE$_n$, and 113\% of the separation energy ratio $S_{\text{PE}_3}/S_{\text{PE}_2}$. Thus, our calculation quantitatively reproduces binding energies and qualitatively reproduces separation energies. These results support the existence of polyexcitons which were hitherto suggested only through experimental photoemission measurements. 

We also obtained bound states of fundamental excitonic complexes like excitons, trions, biexcitons, and charged biexcitons, although the presence of trion$^\pm$ and CBE$^\pm$ is not yet experimentally verified in diamond. Our analysis of separation energies indicates that peak positions of exciton decays in trion$^\pm$ and CBE$^\pm$ mingle with those of biexcitons and triexcitons respectively, in the photoemission spectrum. It would be possible to identify these charged species experimentally in photoluminescence spectrum. This can be done by applying a gate voltage to inject excess carriers in a doped environment so that extra electrons or holes are captured by excitons and form trions\cite{TrionVg1}$^,$\cite{TrionVg2}.

\begin{acknowledgments}
The authors acknowledge financial support from Professional development Consortium Computational Material Scientists (PCoMS). Our work is partially supported by MEXT Elements Strategy Initiative to Form Core Research Center in Japan. We also thank Prof. Maksym for fruitful discussions and proofreading. 
\end{acknowledgments}

\appendix
\section{Binding energies and separation energies of excitonic complexes}
Here we show the value of the total binding energy $E_{\text{PE}_n}$ and separation energy $S_{\text{PE}_n}$ for every possible inequivalent bound state. The columns "electron" and "hole" show the subscript of the valley or band. $(00k_c)$ means the valley on $k_z$ axis and $yz$ means the $\Gamma_{yz}$ band, for instance. There are 6 valleys ($\Delta(\pm k_c 00)$, $\Delta(0\pm k_c0)$, $\Delta(00\pm k_c)$) and 3 bands ($\Gamma_{yz}$, $\Gamma_{zx}$, $\Gamma_{xy}$) in diamond. We regard two combinations as equivalent if they can be transformed into each other by rotation and inversion. The factor $g$ is the degeneracy of each state.
\begin{description}
\item[Exciton]\mbox{}
\begin{longtable}{|c|c|c|c|c|c|c|c|}\hline 
electron&hole&$E_{\text{PE}_n}\text{[meV]}$ & $S_{\text{PE}_n}\text{[meV]}$&$g$\\ \hline\hline
\small{$(00k_c)$}&$xy$& $-7.181\times 10^1$& -&6 \\ \hline
\small{$(00k_c)$} &$yz$& $-6.774\times 10^1$& -&12 \\ \hline
\multicolumn{2}{|c|}{Average}&$- 6.909\times 10^1$& - &\\ \hline
\end{longtable}
\item[Trion$^+$]\mbox{}
\begin{longtable}{|c|c|c|c|c|c|}\hline 
electron&\multicolumn{2}{|c|}{hole}&$E_{\text{PE}_n}\text{[meV]}$ & $S_{\text{PE}_n}\text{[meV]}$&$g$\\ \hline\hline
\small{$(00k_c)$}& $yz$& $yz$&$7.113\times 10^1$& $3.39$&12 \\ \hline
\small{$(00k_c)$}& $yz$& $xy$&$7.345\times 10^1$& $1.64$&12 \\ \hline
\small{$(00k_c)$}& $yz$& $zx$&$7.110\times 10^1$& $3.36$&6 \\ \hline
\small{$(00k_c)$}& $xy$& $xy$&$7.505\times 10^1$& $3.24$&6 \\ \hline
\multicolumn{3}{|c|}{Average}& $7.255\times 10^1$ & $2.78$&\\ \hline
\end{longtable}
\item[Trion$^-$]\mbox{}
\begin{longtable}{|c|c|c|c|c|c|}\hline 
\multicolumn{2}{|c|}{electron}&hole&$E_{\text{PE}_n}\text{[meV]}$ & $S_{\text{PE}_n}\text{[meV]}$&$g$\\ \hline\hline
\small{$(00k_c)$}&\small{$(00k_c)$}& $yz$&$7.138\times 10^1$& $4.09$&12 \\ \hline
\small{$(00k_c)$}&\small{$(00k_c)$}& $xy$&$7.564\times 10^1$& $3.83$&6 \\ \hline
\small{$(00k_c)$}&\small{$(00-k_c)$}& $yz$&$7.183\times 10^1$& $4.09$&6 \\ \hline
\small{$(00k_c)$}&\small{$(00-k_c)$}& $xy$&$7.564\times 10^1$& $3.83$&3 \\ \hline
\small{$(00k_c)$}&\small{$(0k_c0)$}& $yz$&$7.158\times 10^1$& $3.84$&12 \\ \hline
\small{$(00k_c)$}&\small{$(0k_c0)$}& $xy$&$7.386\times 10^1$& $2.05$&24 \\ \hline
\multicolumn{3}{|c|}{Average}& $7.310\times 10^1$ & $3.23$ &\\ \hline
\end{longtable}
\item[Biexciton]\mbox{}
\begin{longtable}{|c|c|c|c|c|c|c|c|}\hline 
\multicolumn{2}{|c|}{electron}&\multicolumn{2}{|c|}{hole}&$E_{\text{PE}_n}\text{[meV]}$ & $S_{\text{PE}_n}\text{[meV]}$&$g$\\ \hline\hline
\small{$(00k_c)$} & \small{$(00k_c)$}& $yz$& $yz$& $1.412\times 10^2$ &$5.73$&12 \\ \hline
\small{$(00k_c)$} & \small{$(00k_c)$}& $yz$& $zx$& $1.409\times 10^2$ &$5.40$&6 \\ \hline
\small{$(00k_c)$} & \small{$(00k_c)$}& $yz$& $xy$& $1.446\times 10^2$ &$5.04$&12 \\ \hline
\small{$(00k_c)$} & \small{$(00k_c)$}& $xy$& $xy$& $1.482\times 10^2$ &$4.60$&6\\ \hline
\small{$(00k_c)$} & \small{$(00-k_c)$}& $yz$& $yz$&$1.410\times 10^2$ &$5.49$&6 \\ \hline
\small{$(00k_c)$} & \small{$(00-k_c)$}& $yz$& $zx$&$1.411\times 10^2$ &$5.59$&3 \\ \hline
\small{$(00k_c)$} & \small{$(00-k_c)$}& $yz$& $xy$&$1.446\times 10^2$ &$5.01$&6 \\ \hline
\small{$(00k_c)$} & \small{$(00-k_c)$}& $xy$& $xy$&$1.479\times 10^2$ &$4.24$&3 \\ \hline
\small{$(00k_c)$} & \small{$(0k_c0)$}& $yz$& $yz$&$1.407\times 10^2$ &$5.20$&12 \\ \hline
\small{$(00k_c)$} & \small{$(0k_c0)$}& $yz$& $zx$&$1.427\times 10^2$ &$3.13$&24 \\ \hline
\small{$(00k_c)$} & \small{$(0k_c0)$}& $xy$& $xy$&$1.443\times 10^2$ &$4.74$&24 \\ \hline
\small{$(00k_c)$} & \small{$(0k_c0)$}& $xy$& $zx$&$1.450\times 10^2$ &$1.40$&12 \\ \hline
\multicolumn{4}{|c|}{Average}& $1.433\times 10^2$ & $4.36$ &\\ \hline
\end{longtable}
\item[CBE$^+$]\mbox{}
\begin{longtable}{|c|c|c|c|c|c|c|c|}\hline 
%\begin{tabular}{|c|c|c|c|c|c|c|}\hline 
\multicolumn{2}{|c|}{electron}&\multicolumn{3}{|c|}{hole}&$E_{\text{PE}_n}\text{[meV]}$ & $S_{\text{PE}_n}\text{[meV]}$&$g$\\ \hline\hline
\small{$(00k_c)$}& \small{$(00k_c)$}& $\small{yz}$&\small{ $yz$}&\small{ $yz$}&\small{unbound} &\small{-}&-\\ \hline
\small{$(00k_c)$}& \small{$(00k_c)$}& $\small{yz}$&\small{ $yz$}&\small{ $zx$}&\small{$1.521\times 10^2$} &\small{$1.09\times 10^1$}&12\\ \hline
\small{$(00k_c)$}& \small{$(00k_c)$}& $\small{yz}$&\small{ $yz$}&\small{ $xy$}&\small{$1.534\times 10^2$} &\small{$8.77$}&12\\ \hline
\small{$(00k_c)$}& \small{$(00k_c)$}& $\small{yz}$&\small{ $zx$}&\small{ $xy$}&\small{$1.537\times 10^2$} &\small{$9.07$}&6\\ \hline
\small{$(00k_c)$}& \small{$(00k_c)$}& $\small{yz}$&\small{ $xy$}&\small{ $xy$}&\small{$1.551\times 10^2$} &\small{$6.86$}&12\\ \hline
\small{$(00k_c)$}& \small{$(00k_c)$}& $\small{xy}$&\small{ $xy$}&\small{ $xy$}&\small{unbound} &\small{-}&-\\ \hline
\small{$(00k_c)$}& \scriptsize{$(00$ -$k_c)$}& $\small{yz}$&\small{ $yz$}&\small{ $yz$}&\small{unbound} &\small{-}&-\\ \hline
\small{$(00k_c)$}& \scriptsize{$(00$ -$k_c)$}& $\small{yz}$&\small{ $yz$}&\small{ $zx$}&\small{$1.516\times 10^2$} &\small{$1.06\times 10^1$}&6\\ \hline
\small{$(00k_c)$}& \scriptsize{$(00$ -$k_c)$}& $\small{yz}$&\small{ $yz$}&\small{ $xy$}&\small{$1.539\times 10^2$} &\small{$9.30$}&6\\ \hline
\small{$(00k_c)$}& \scriptsize{$(00$ -$k_c)$}& $\small{yz}$&\small{ $zx$}&\small{ $xy$}&\small{$1.540\times 10^2$} &\small{$9.44$}&3\\ \hline
\small{$(00k_c)$}& \scriptsize{$(00$ -$k_c)$}& $\small{yz}$&\small{ $xy$}&\small{ $xy$}&\small{$1.555\times 10^2$} &\small{$7.62\times 10^1$}&6\\ \hline
\small{$(00k_c)$}& \scriptsize{$(00$ -$k_c)$}& $\small{xy}$&\small{ $xy$}&\small{ $xy$}&\small{unbound} &\small{-}&-\\ \hline
\small{$(00k_c)$}& \small{$(0k_c0)$}& $\small{yz}$&\small{ $yz$}&\small{ $yz$}&\small{unbound} &\small{-}&-\\ \hline
\small{$(00k_c)$}& \small{$(0k_c0)$}& $\small{yz}$&\small{ $yz$}&\small{ $zx$}&\small{$1.523\times 10^2$} &\small{$9.61$}&24\\ \hline
\small{$(00k_c)$}& \small{$(0k_c0)$}& $\small{yz}$&\small{ $zx$}&\small{ $zx$}&\small{$1.533\times 10^2$} &\small{$8.97$}&24\\ \hline
\small{$(00k_c)$}& \small{$(0k_c0)$}& $\small{yz}$&\small{ $zx$}&\small{ $xy$}&\small{$1.539\times 10^2$} &\small{$8.85$}&12\\ \hline
\small{$(00k_c)$}& \small{$(0k_c0)$}& $\small{xy}$&\small{ $xy$}&\small{ $zx$}&\small{$1.546\times 10^2$} &\small{$9.60$}&24\\ \hline
\small{$(00k_c)$}& \small{$(0k_c0)$}& $\small{xy}$&\small{ $xy$}&\small{ $xy$}&\small{unbound} &\small{-}&-\\ \hline
\multicolumn{5}{|c|}{Average}& $1.535\times 10^2$ & $9.31$ &\\ \hline
%\end{tabular}
\end{longtable}
\item[CBE$^-$]\mbox{}
\begin{longtable}{|c|c|c|c|c|c|c|c|}\hline 
\multicolumn{3}{|c|}{electron}&\multicolumn{2}{|c|}{hole}&$E_{\text{PE}_n}\text{[meV]}$ & $S_{\text{PE}_n}\text{[meV]}$&$g$\\ \hline\hline
\footnotesize{$(00k_c)$}& \footnotesize{$(00k_c)$}&\footnotesize{$(00k_c)$}& \footnotesize{$yz$}&\footnotesize{$yz$}&\footnotesize{unbound}&\footnotesize{-}&-\\ \hline
\footnotesize{$(00k_c)$}& \footnotesize{$(00k_c)$}&\footnotesize{$(0k_c0)$}& \footnotesize{$yz$}&\footnotesize{$yz$}&\footnotesize{$1.513\times 10^2$}&\footnotesize{$1.01\times 10^1$}&24\\ \hline
\footnotesize{$(00k_c)$}& \footnotesize{$(00k_c)$}&\footnotesize{$(k_c00)$}& \footnotesize{$yz$}&\footnotesize{$yz$}&\footnotesize{$1.537\times 10^2$}&\footnotesize{$9.11$}&24\\ \hline
\footnotesize{$(00k_c)$}& \footnotesize{$(00k_c)$}&\footnotesize{$(00-k_c)$}& \footnotesize{$yz$}&\footnotesize{$yz$}&\footnotesize{$1.524\times 10^2$}&\footnotesize{$1.11\times 10^1$}&12\\ \hline
\footnotesize{$(00k_c)$}& \footnotesize{$(00k_c)$}&\footnotesize{$(00k_c)$}& \footnotesize{$yz$}&\footnotesize{$zx$}&\footnotesize{unbound}&\footnotesize{-}&-\\ \hline
\footnotesize{$(00k_c)$}& \footnotesize{$(00k_c)$}&\footnotesize{$(k_c00)$}& \footnotesize{$yz$}&\footnotesize{$zx$}&\footnotesize{$1.527\times 10^2$}&\footnotesize{$1.18\times 10^1$}&12\\ \hline
\footnotesize{$(00k_c)$}& \footnotesize{$(00k_c)$}&\footnotesize{$(00-k_c)$}& \footnotesize{$yz$}&\footnotesize{$zx$}&\footnotesize{$1.518\times 10^2$}&\footnotesize{$1.07\times 10^2$}&6\\ \hline
\footnotesize{$(00k_c)$}& \footnotesize{$(00k_c)$}&\footnotesize{$(00k_c)$}& \footnotesize{$yz$}&\footnotesize{$xy$}&\footnotesize{unbound}&\footnotesize{-}&-\\ \hline
\footnotesize{$(00k_c)$}& \footnotesize{$(00k_c)$}&\footnotesize{$(00-k_c)$}& \footnotesize{$yz$}&\footnotesize{$xy$}&\footnotesize{$1.547\times 10^2$}&\footnotesize{$1.01\times 10^2$}&12\\ \hline
\footnotesize{$(00k_c)$}& \footnotesize{$(00k_c)$}&\footnotesize{$(k_c00)$}& \footnotesize{$yz$}&\footnotesize{$xy$}&\footnotesize{$1.549\times 10^2$}&\footnotesize{$9.91$}&24\\ \hline
\footnotesize{$(00k_c)$}& \footnotesize{$(00k_c)$}&\footnotesize{$(0k_c0)$}& \footnotesize{$yz$}&\footnotesize{$xy$}&\footnotesize{$1.538\times 10^2$}&\footnotesize{$9.18$}&24\\ \hline
\footnotesize{$(00k_c)$}& \footnotesize{$(00k_c)$}&\footnotesize{$(00k_c)$}& \footnotesize{$xy$}&\footnotesize{$xy$}&\footnotesize{unbound}&\footnotesize{-}&-\\ \hline
\footnotesize{$(00k_c)$}& \footnotesize{$(00k_c)$}&\footnotesize{$(00-k_c)$}& \footnotesize{$xy$}&\footnotesize{$xy$}&\footnotesize{$1.587\times 10^2$}&\footnotesize{$1.05\times 10^1$}&6\\ \hline
\footnotesize{$(00k_c)$}& \footnotesize{$(00k_c)$}&\footnotesize{$(k_c00)$}& \footnotesize{$xy$}&\footnotesize{$xy$}&\footnotesize{$1.552\times 10^2$}&\footnotesize{$6.98$}&24\\ \hline
\footnotesize{$(00k_c)$}& \footnotesize{$(00$ -$k_c)$}&\footnotesize{$(k_c00)$}& \footnotesize{$yz$}&\footnotesize{$yz$}&\footnotesize{$1.540\times 10^2$}&\footnotesize{$9.46$}&12\\ \hline
\footnotesize{$(00k_c)$}& \footnotesize{$(00$ -$k_c)$}&\footnotesize{$(0k_c0)$}& \footnotesize{$yz$}&\footnotesize{$yz$}&\footnotesize{$1.516\times 10^2$}&\footnotesize{$1.07\times 10^1$}&12\\ \hline
\footnotesize{$(00k_c)$}& \footnotesize{$(00$ -$k_c)$}&\footnotesize{$(k_c00)$}& \footnotesize{$yz$}&\footnotesize{$zx$}&\footnotesize{$1.530\times 10^2$}&\footnotesize{$1.03\times 10^1$}&12\\ \hline
\footnotesize{$(00k_c)$}& \footnotesize{$(00$ -$k_c)$}&\footnotesize{$(k_c00)$}& \footnotesize{$yz$}&\footnotesize{$xy$}&\footnotesize{$1.554\times 10^2$}&\footnotesize{$1.04\times 10^1$}&12\\ \hline
\footnotesize{$(00k_c)$}& \footnotesize{$(00$ -$k_c)$}&\footnotesize{$(k_c00)$}& \footnotesize{$xy$}&\footnotesize{$xy$}&\footnotesize{$1.559\times 10^2$}&\footnotesize{$7.70$}&12\\ \hline
\footnotesize{$(00k_c)$}& \footnotesize{$(0k_c0)$}&\footnotesize{$(0-k_c0)$}& \footnotesize{$yz$}&\footnotesize{$zx$}&\footnotesize{$1.540\times 10^2$}&\footnotesize{$9.49$}&12\\ \hline
\footnotesize{$(00k_c)$}& \footnotesize{$(0k_c0)$}&\footnotesize{$(k_c00)$}& \footnotesize{$xy$}&\footnotesize{$xy$}&\footnotesize{$1.539\times 10^2$}&\footnotesize{$9.57$}&24\\ \hline
\footnotesize{$(00k_c)$}& \footnotesize{$(0k_c0)$}&\footnotesize{$(k_c00)$}& \footnotesize{$xy$}&\footnotesize{$zx$}&\footnotesize{$1.540\times 10^2$}&\footnotesize{$9.00$}&48\\ \hline
\multicolumn{5}{|c|}{Average}& $1.539\times 10^2$ & $9.51$ &\\ \hline
\end{longtable}
\item[Triexciton]\mbox{}
\begin{longtable}{|c|c|c|c|c|c|c|c|c|}\hline 
\multicolumn{3}{|c|}{electron}&\multicolumn{3}{|c|}{hole}& \footnotesize{$E_{\text{PE}_n}\text{[meV]}$ }& \footnotesize{$S_{\text{PE}_n}\text{[meV]}$}&$g$\\ \hline\hline
\scriptsize{$(00k_c)$}&\scriptsize{$(00k_c)$}& \scriptsize{$(00k_c)$}      & \scriptsize{$yz$}&\scriptsize{ $yz$}&\scriptsize{ $yz$}&\scriptsize{unbound}&\scriptsize{-}&-\\ \hline
\scriptsize{$(00k_c)$}&\scriptsize{$(00k_c)$}& \scriptsize{$(00k_c)$}      & \scriptsize{$yz$}&\scriptsize{ $yz$}&\scriptsize{ $zx$}&\scriptsize{unbound}&\scriptsize{-}&-\\ \hline
\scriptsize{$(00k_c)$}&\scriptsize{$(00k_c)$}& \scriptsize{$(00k_c)$}      & \scriptsize{$yz$}&\scriptsize{ $yz$}&\scriptsize{ $xy$}&\scriptsize{unbound}&\scriptsize{-}&-\\ \hline
\scriptsize{$(00k_c)$}&\scriptsize{$(00k_c)$}& \scriptsize{$(0k_c0)$}      & \scriptsize{$yz$}&\scriptsize{ $yz$}&\scriptsize{ $yz$}&\scriptsize{unbound}&\scriptsize{-}&-\\ \hline
\scriptsize{$(00k_c)$}&\scriptsize{$(00k_c)$}& \scriptsize{$(0k_c0)$}      & \scriptsize{$yz$}&\scriptsize{ $yz$}&\scriptsize{ $zx$}&\scriptsize{$2.229\times 10^2$}&\scriptsize{$9.91$}&24\\ \hline
\scriptsize{$(00k_c)$}&\scriptsize{$(00k_c)$}& \scriptsize{$(0k_c0)$}      & \scriptsize{$yz$}&\scriptsize{ $yz$}&\scriptsize{ $xy$}&\scriptsize{$2.238\times 10^2$}&\scriptsize{$1.13\times 10^1$}&24\\ \hline
\scriptsize{$(00k_c)$}&\scriptsize{$(00k_c)$}& \scriptsize{$(k_c00)$}	      & \scriptsize{$zy$}&\scriptsize{ $yz$}&\scriptsize{ $yz$}&\scriptsize{unbound}&\scriptsize{-}&-\\ \hline
\scriptsize{$(00k_c)$}&\scriptsize{$(00k_c)$}& \scriptsize{$(k_c00)$}      & \scriptsize{$yz$}&\scriptsize{ $yz$}&\scriptsize{ $zx$}&\scriptsize{$2.259\times 10^2$}&\scriptsize{$1.32\times 10^1$}&24\\ \hline
\scriptsize{$(00k_c)$}&\scriptsize{$(00k_c)$}& \scriptsize{$(k_c00)$}      & \scriptsize{$yz$}&\scriptsize{ $yz$}&\scriptsize{ $xy$}&\scriptsize{$2.278\times 10^2$}&\scriptsize{$1.14\times 10^1$}&24\\ \hline
\scriptsize{$(00k_c)$}&\scriptsize{$(00k_c)$}& \scriptsize{$(00-k_c)$}& \scriptsize{$yz$}&\scriptsize{ $yz$}&\scriptsize{ $yz$}&\scriptsize{unbound}&\scriptsize{-}&-\\ \hline
\scriptsize{$(00k_c)$}&\scriptsize{$(00k_c)$}& \scriptsize{$(00-k_c)$}& \scriptsize{$yz$}&\scriptsize{ $yz$}&\scriptsize{ $zx$}&\scriptsize{$2.223\times 10^2$}&\scriptsize{$1.34\times 10^1$}&12\\ \hline
\scriptsize{$(00k_c)$}&\scriptsize{$(00k_c)$}& \scriptsize{$(00-k_c)$}& \scriptsize{$yz$}&\scriptsize{ $yz$}&\scriptsize{ $xy$}&\scriptsize{$2.249\times 10^2$}&\scriptsize{$1.19\times 10^1$}&12\\ \hline
\scriptsize{$(00k_c)$}&\scriptsize{$(00k_c)$}& \scriptsize{$(00k_c)$}      & \scriptsize{$zy$}&\scriptsize{ $zx$}&\scriptsize{ $yz$}&\scriptsize{unbound}&\scriptsize{-}&-\\ \hline
\scriptsize{$(00k_c)$}&\scriptsize{$(00k_c)$}& \scriptsize{$(00k_c)$}      & \scriptsize{$yz$}&\scriptsize{ $zx$}&\scriptsize{ $xy$}&\scriptsize{unbound}&\scriptsize{-}&-\\ \hline
\scriptsize{$(00k_c)$}&\scriptsize{$(00k_c)$}& \scriptsize{$(00-k_c)$}      & \scriptsize{$yz$}&\scriptsize{ $zx$}&\scriptsize{ $xy$}&\scriptsize{$2.249\times 10^2$}&\scriptsize{$1.20\times 10^1$}&6\\ \hline
\scriptsize{$(00k_c)$}&\scriptsize{$(00k_c)$}& \scriptsize{$(00k_c)$}& \scriptsize{$yz$}&\scriptsize{ $xy$}&\scriptsize{ $xy$}&\scriptsize{unbound}&\scriptsize{-}&-\\ \hline
\scriptsize{$(00k_c)$}&\scriptsize{$(00k_c)$}& \scriptsize{$(00-k_c)$}& \scriptsize{$yz$}&\scriptsize{ $xy$}&\scriptsize{ $xy$}&\scriptsize{$2.292\times 10^2$}&\scriptsize{$1.28\times 10^1$}&12\\ \hline
\scriptsize{$(00k_c)$}&\scriptsize{$(00k_c)$}& \scriptsize{$(k_c00)$}& \scriptsize{$yz$}&\scriptsize{ $xy$}&\scriptsize{ $xy$}&\scriptsize{$2.266\times 10^2$}&\scriptsize{$6.56$}&24\\ \hline
\scriptsize{$(00k_c)$}&\scriptsize{$(00k_c)$}& \scriptsize{$(0k_c0)$}& \scriptsize{$yz$}&\scriptsize{ $xy$}&\scriptsize{ $zx$}&\scriptsize{$2.254\times 10^2$}&\scriptsize{$9.01$}&24\\ \hline
\scriptsize{$(00k_c)$}&\scriptsize{$(00k_c)$}& \scriptsize{$(00k_c)$}& \scriptsize{$xy$}&\scriptsize{ $xy$}&\scriptsize{ $yx$}&\scriptsize{unbound}&\scriptsize{-}&-\\ \hline
\scriptsize{$(00k_c)$}&\scriptsize{$(00k_c)$}& \scriptsize{$(00-k_c)$}& \scriptsize{$xy$}&\scriptsize{ $xy$}&\scriptsize{ $xy$}&\scriptsize{unbound}&\scriptsize{-}&-\\ \hline
\scriptsize{$(00k_c)$}&\scriptsize{$(00k_c)$}& \scriptsize{$(k_c00)$}& \scriptsize{$xy$}&\scriptsize{ $xy$}&\scriptsize{ $zx$}&\scriptsize{$2.256\times 10^2$}&\scriptsize{$9.66$}&24\\ \hline
\scriptsize{$(00k_c)$}&\scriptsize{$(00k_c)$}& \scriptsize{$(k_c00)$}& \scriptsize{$xy$}&\scriptsize{ $xy$}&\scriptsize{ $xy$}&\scriptsize{unbound}&\scriptsize{-}&-\\ \hline
\scriptsize{$(00k_c)$}&\scriptsize{$(00$ -$k_c)$}& \scriptsize{$(k_c00)$}& \scriptsize{$yz$}&\scriptsize{ $yz$}&\scriptsize{ $yz$}&\scriptsize{unbound}&\scriptsize{-}&-\\ \hline
\scriptsize{$(00k_c)$}&\scriptsize{$(00$ -$k_c)$}& \scriptsize{$(k_c00)$}& \scriptsize{$yz$}&\scriptsize{ $yz$}&\scriptsize{ $zx$}&\scriptsize{$2.245\times 10^2$}&\scriptsize{$1.17\times 10^1$}&12\\ \hline
\scriptsize{$(00k_c)$}&\scriptsize{$(00$ -$k_c)$}& \scriptsize{$(k_c00)$}& \scriptsize{$yz$}&\scriptsize{ $yz$}&\scriptsize{ $xy$}&\scriptsize{$2.265\times 10^2$}&\scriptsize{$1.02\times 10^1$}&12\\ \hline
\scriptsize{$(00k_c)$}&\scriptsize{$(00$ -$k_c)$}& \scriptsize{$(0k_c0)$}& \scriptsize{$yz$}&\scriptsize{ $yz$}&\scriptsize{ $yz$}&\scriptsize{unbound}&\scriptsize{-}&-\\ \hline
\scriptsize{$(00k_c)$}&\scriptsize{$(00$ -$k_c)$}& \scriptsize{$(0k_c0)$}& \scriptsize{$yz$}&\scriptsize{ $yz$}&\scriptsize{ $zx$}&\scriptsize{$2.237\times 10^2$}&\scriptsize{$1.09\times 10^1$}&12\\ \hline
\scriptsize{$(00k_c)$}&\scriptsize{$(00$ -$k_c)$}& \scriptsize{$(0k_c0)$}& \scriptsize{$yz$}&\scriptsize{ $yz$}&\scriptsize{ $xy$}&\scriptsize{$2.243\times 10^2$}&\scriptsize{$1.18\times 10^1$}&12\\ \hline
\scriptsize{$(00k_c)$}&\scriptsize{$(00$ -$k_c)$}& \scriptsize{$(k_c00)$}& \scriptsize{$yz$}&\scriptsize{ $zx$}&\scriptsize{ $xy$}&\scriptsize{$2.257\times 10^2$}&\scriptsize{$9.33$}&12\\ \hline
\scriptsize{$(00k_c)$}&\scriptsize{$(00$ -$k_c)$}& \scriptsize{$(k_c00)$}& \scriptsize{$yz$}&\scriptsize{ $xy$}&\scriptsize{ $xy$}&\scriptsize{$2.275\times 10^2$}&\scriptsize{$7.87$}&12\\ \hline
\scriptsize{$(00k_c)$}&\scriptsize{$(00$ -$k_c)$}& \scriptsize{$(k_c00)$}& \scriptsize{$xy$}&\scriptsize{ $xy$}&\scriptsize{ $zx$}&\scriptsize{$2.265\times 10^2$}&\scriptsize{$1.09\times 10^1$}&12\\ \hline
\scriptsize{$(00k_c)$}&\scriptsize{$(00$ -$k_c)$}& \scriptsize{$(k_c00)$}& \scriptsize{$xy$}&\scriptsize{ $xy$}&\scriptsize{ $xy$}&\scriptsize{unbound}&\scriptsize{-}&-\\ \hline
\scriptsize{$(00k_c)$}&\scriptsize{$(0k_c0)$}& \scriptsize{$(k_c00)$}& \scriptsize{$xy$}&\scriptsize{ $xy$}&\scriptsize{ $xy$}&\scriptsize{unbound}&\scriptsize{-}&-\\ \hline
\scriptsize{$(00k_c)$}&\scriptsize{$(0k_c0)$}& \scriptsize{$(k_c00)$}& \scriptsize{$xy$}&\scriptsize{ $xy$}&\scriptsize{ $yz$}&\scriptsize{$2.246\times 10^2$}&\scriptsize{$8.55$}&48\\ \hline
\scriptsize{$(00k_c)$}&\scriptsize{$(0k_c0)$}& \scriptsize{$(k_c00)$}& \scriptsize{$xy$}&\scriptsize{ $zx$}&\scriptsize{ $yz$}&\scriptsize{$2.255\times 10^2$}&\scriptsize{$8.62$}&8\\ \hline
\multicolumn{6}{|c|}{Average}& \scriptsize{$2.253\times 10^2$ }& \scriptsize{$1.02\times 10^1$} &\\ \hline
\end{longtable}
\end{description}

\section{Matrix elements of Hamiltonian}
Here we give the matrix elements of the Gram matrix and the Hamiltonian with multiple valley and band in anisotropic systems. The details of derivations of the matrix elements for isotropic systems are in the text book of Suzuki and Varga\cite{CGtext}. The book does not treat anisotropic systems, but the derivations described there can be generalized to anisotropic systems as follows. We assume the following type of basis.
\begin{eqnarray}
|LM, \bm{v}, A\rangle = f_{LM}(\bm{r})\cdot \chi_{sm_s} \cdot \prod_{i = 1}^{N_e}|\Delta_i\rangle\prod_{i = 1}^{N_h}|\Gamma_i\rangle 
\end{eqnarray}
The envelope function $f_{LM}(\bm{r})$ is given by a product of a solid spherical harmonic, a Gaussian, and a plane wave part.
\begin{eqnarray}
&&f_{LM}(\bm{r}) =\nonumber\\
&& |\bm{v}|^{L}Y_{LM}(\hat{\bm{v}})\cdot \exp\{-\frac{1}{2}\mathrm{xAx}\} \prod_{i}^{N_e}\exp\{i\bm{k_{\Delta_i}\cdot r_i}\}
\end{eqnarray}
The matrix elements of the Gram matrix is given by
\begin{eqnarray}
&&\langle L'M',\bm{v}', A'|LM,\bm{v}, A\rangle\nonumber \\
&&= \frac{(2L+1)!!}{4\pi}\left\{\frac{(2\pi)^{N-1}}{\det B}\right\}^{\frac{3}{2}}\rho^L\delta_{L'L}\delta_{M'M}\nonumber\\
&&\times \langle \chi_{s'm'}|\chi_{sm}\rangle \prod_{i}^{N_e}\delta_{\Delta'_i, \Delta_i}\prod_{i}^{N_h}\delta_{\Gamma'_i, \Gamma_i},
\end{eqnarray}
\begin{eqnarray}
\rho = \sum_{i,j}^{N-1}u'_i (B^{-1})_{ij}u_j
\end{eqnarray}
where $B = A'+A$ and $u_i$ is a coefficient of the global vector $\bm{v} = \sum_{i}^{N-1}u_i\mathrm{x}_i$. Next, we give matrix elements of the anisotropic kinetic energy of the electron:
\begin{eqnarray}
&&\langle L'M',\bm{v}', A'|\sum_{i=1}^{N_e}\sum_{\gamma}t_{e,i}^{(\gamma)} \cdot \hat{\tau}_{\gamma \gamma,i}|LM,\bm{v}, A\rangle\nonumber \\
&&=\frac{1}{2}(B_{0L}B_{0L'})^{-1}\left\{ \frac{(2\pi)}{\det B}\right\}^{\frac{3}{2}}L!L'!\nonumber \\
&&\times \{ g_1(L,M,L',M',R_x,R_y,R_z)\nonumber\\
&&+ g_2(L,M,L',M',P_x,P_y,P_z)\nonumber\\
&& + g_2(L',M',L,M,P'_x,P'_y,P'_z)\nonumber\\
&& + g_3(L,M,L',M',Q_x,Q_y,Q_z) \}\nonumber\\
&& \times \langle \chi_{s'm'}|\chi_{sm}\rangle,
\end{eqnarray}
\begin{eqnarray}
B_{nl} = \frac{4\pi(2n+l)!}{2^nn!(2n+2l+1)!!}.
\end{eqnarray}
The functions $g_1, g_2, and g_3$ are defined as follows:
\begin{eqnarray}
&&g_1(L,M,L',M',R_x,R_y,R_z) \nonumber \\
&&= \frac{1}{L!}(\sum_{\mu = x,y,x}R_\mu)\rho^LB_{0L}\delta_{LL'}\delta_{MM'},
\end{eqnarray}
\begin{eqnarray}
&&R_\mu = \mathrm{Tr} \{A'\Lambda_{\mu}AB^{-1}\} \\
&&\Lambda_{\mu,jk} = \prod_{i= 1}\delta_{\Delta'_i\Delta_i}\sum_{i = 1}\frac{1}{m^{(\Delta_i)}_{i,\mu}}U_{ji}U_{ki},
\end{eqnarray}
\begin{eqnarray}
&&g_2(L,M,L',M',P_x,P_y,P_z) \nonumber \\
&&= \{ \frac{1}{3c_{20}}\left(P_{z}-\frac{P_x+P_y}{2}\right)C_{20,LM}^{L'M}\delta_{MM'}\nonumber \\
&&+\frac{1}{4c_{22}}(P_{x}-P_{y})(C_{2-2,LM}^{L'M-2}\delta_{M'M-2}+C_{22,LM}^{L'M+2}\delta_{M'M+2})\}\nonumber \\
&&\times B_{0L}D_{2LL'}\frac{1}{L'!}\rho^{L'}\delta_{L'L-2},
\end{eqnarray}
\begin{eqnarray}
P_{\mu} = -\sum_{ij}^{N_{all}-1}u_i\{B^{-1}A'\Lambda_{\mu}A'B^{-1}\}_{ij}u_j,
\end{eqnarray}
\begin{eqnarray}
P'_{\mu} = -\sum_{ij}^{N_{all}-1}u'_i\{B^{-1}A\Lambda_{\mu}AB^{-1}\}_{ij}u'_j.
\end{eqnarray}
Here, $C_{l_1m_1,l_2m_2}^{l_3m_3}$ is a Clebsch-Gordan coefficient, $N_{all} = N_e+N_h$, and 
\begin{eqnarray}
c_{20} = \frac{1}{4}\left(\frac{5}{\pi}\right)^{1/2}, \; c_{22} = \frac{1}{4}\left(\frac{15}{2\pi}\right)^{1/2},\\
D_{l_3l_1l_2} = \left\{\frac{(2l_1+1)(2l_2+1)}{4\pi(2l_3+1)}\right\}^{1/2}C_{l_10l_20}^{l_30},
\end{eqnarray}
\begin{eqnarray}
&&g_3(L,M,L',M',Q_x,Q_y,Q_z)\nonumber\\
&&=\frac{1}{c_{10}^2}\{(Q_x-Q_y)[C_{LM,1-1}^{L-1M-1}C_{L-1M'+1,1-1}^{LM'}\delta_{M-1M'+1}\nonumber \\
&&+C_{LM,1+1}^{L-1M+1}C_{L-1M'-1,1+1}^{LM'}\delta_{M+1M'-1}]\nonumber \\
&&-(Q_x+Q_y)[C_{LM,1-1}^{L-1,M-1}C_{L-1M'-1,1+1}^{LM'}\nonumber \\
&&+C_{LM,1+1}^{L-1M+1}C_{L-1M'+1,1-1}^{LM'}]\delta_{MM'}\nonumber \\
&&+2Q_zC_{LM,10}^{L-1M}C_{L-1M',10}^{LM'}\delta_{MM'}\}\nonumber \\
&&\times D_{L-1L1}D_{LL-11}\delta_{LL'}B_{0L-1}\rho^{L-1},
\end{eqnarray}
\begin{eqnarray}
Q_\mu = 2\sum_{ij}^{N_{all}-1}u'_i\{B^{-1}A\Lambda_{\mu}A'B^{-1}\}_{ij}u_j, \;c_{10} = \left(\frac{3}{2\pi}\right)^{1/2}.\nonumber \\
\end{eqnarray}
These expressions are applied to the matrix elements of hole kinetic energy. Next we give the matrix elements of inter-band coupling.
\begin{eqnarray}
&&\langle L'M',\bm{v}', A'| \sum_{i=1}^{N_h} t_{h,i}^{(\Gamma_{xy} \Gamma_{yz})}\cdot \hat{\tau}_{\Gamma_{xy} \Gamma_{yz},i}|LM,\bm{v}, A\rangle\nonumber \\
&&= (B_{0L'}B_{0L})^{-1}L'!L!\left\{ \frac{(2\pi)^{N-1}}{\det B}\right\}\nonumber \\
&&\times\{h_{1,xz}(L,M,L',M',S_{xz})\nonumber \\
&&+h_{1,xz}(L',M',L,M,S'_{xz})\nonumber \\
&&+h_{2,xz}(L,M,L',M',T_{xz})\}\langle \chi_{s'm'}|\chi_{sm}\rangle,
\end{eqnarray}
\begin{eqnarray}
&&h_{1,xz}(L,M,L',M',S_{xz})\nonumber\\
&&=\frac{S_{xz}}{4c_{22}}D_{L2L'}(C_{LM,2-1}^{L'M-1}\delta_{M',M-1}-C_{LM,21}^{L'M+1}\delta_{M',M+1})\nonumber \\
&&\times B_{0L}'\frac{1}{L'!}\rho^{L'}\delta_{L',L-2},\nonumber \\
\end{eqnarray}
\begin{eqnarray}
&&h_{2,xz}(L,M,L',M',T_{xz})\nonumber \\
&&= -\frac{T_{xz}}{2^{3/2}c_{11}^2}D_{L-1,1,L}D_{L,1,L-1}B_{0L-1}\frac{1}{(L-1)!}\rho^{L-1}\delta_{L'L}\nonumber \\
&&\times\{ C_{LM,10}^{L-1M}(C_{L-1M,1-1}^{LM'}\delta_{M'M-1}-C_{L-1M,11}^{LM'}\delta_{M'M+1})\nonumber \\
&&+ C_{L-1M',10}^{LM'}(C_{LM,1-1}^{L-1M'}\delta_{M'M-1}-C_{LM,11}^{L-1M'}\delta_{M'M+1})\},\nonumber \\
\end{eqnarray}
\begin{eqnarray}
&&\langle L'M',\bm{v}', A'| \sum_{i=1}^{N_h} t_{h,i}^{(\Gamma_{yz} \Gamma_{zx})}\cdot \hat{\tau}_{\Gamma_{yz} \Gamma_{zx},i}|LM,\bm{v}, A\rangle\nonumber \\
&&=i(B_{0L'}B_{0L})^{-1}L'!L!\left\{ \frac{(2\pi)^{N-1}}{\det B}\right\}\nonumber \\
&&\times\{h_{1,xy}(L,M,L',M',S_{xy})\nonumber \\
&&+h_{1,xy}(L',M',L,M,S'_{xy})\nonumber \\
&&+h_{2,xy}(L,M,L',M',T_{xy})\}\langle \chi_{s'm'}|\chi_{sm}\rangle,
\end{eqnarray}
\begin{eqnarray}
&&h_{1,xy}(L,M,L',M',S_{xy})\nonumber\\
&&=\frac{S_{xy}}{4c_{22}}D_{L2L'}(C_{LM,2-2}^{L'M-2}\delta_{M',M-2}-C_{LM,22}^{L'M+2}\delta_{M',M+2})\nonumber \\
&&\times B_{0L}'\frac{1}{L'!}\rho^{L'}\delta_{L',L-2},\nonumber \\
\end{eqnarray}
\begin{eqnarray}
&&h_{2,xy}(L,M,L',M',T_{xy})\nonumber \\
&&= -\frac{T_{xy}}{2c_{11}^2}D_{L-1,1,L}D_{L,1,L-1}B_{0L-1}\frac{1}{(L-1)!}\rho^{L-1}\delta_{L'L}\nonumber \\
&&\times( C_{LM,1-1}^{L-1M-1}C_{L-1M'+1,1-1}^{L'M'}\delta_{M'+1M-1}\nonumber \\
&&-C_{LM,11}^{L-1M+1}C_{L-1M'-1,11}^{L'M'}\delta_{M'-1M+1}),\nonumber \\
\end{eqnarray}
\begin{eqnarray}
&&\langle L'M',\bm{v}', A'| \sum_{i=1}^{N_h} t_{h,i}^{(\Gamma_{zx} \Gamma_{xy})}\cdot \hat{\tau}_{\Gamma_{zx} \Gamma_{xy},i}|LM,\bm{v}, A\rangle\nonumber \\
&&= i(B_{0L'}B_{0L})^{-1}L'!L!\left\{ \frac{(2\pi)^{N-1}}{\det B}\right\}\nonumber \\
&&\times\{h_{1,yz}(L,M,L',M',S_{yz})\nonumber \\
&&+h_{1,yz}(L',M',L,M,S'_{yz})\nonumber \\
&&+h_{2,yz}(L,M,L',M',T_{yz})\}\langle \chi_{s'm'}|\chi_{sm}\rangle,
\end{eqnarray}
\begin{eqnarray}
&&h_{1,yz}(L,M,L',M',S_{xz})\nonumber\\
&&=\frac{S_{yz}}{4c_{22}}D_{L2L'}(C_{LM,2-1}^{L'M-1}\delta_{M',M-1}+C_{LM,21}^{L'M+1}\delta_{M',M+1})\nonumber \\
&&\times B_{0L}'\frac{1}{L'!}\rho^{L'}\delta_{L',L-2},\nonumber\\
\end{eqnarray}
\begin{eqnarray}
&&h_{2,yz}(L,M,L',M',T_{xz})\nonumber \\
&&= -\frac{T_{yz}}{2^{3/2}c_{11}^2}D_{L-1,1,L}D_{L,1,L-1}B_{0L-1}\frac{1}{(L-1)!}\rho^{L-1}\delta_{L'L}\nonumber \\
&&\times\{ C_{LM,10}^{L-1M}(C_{L-1M,1-1}^{LM'}\delta_{M'M-1}+C_{L-1M,11}^{LM'}\delta_{M'M+1})\nonumber \\
&&+ C_{L-1M',10}^{LM'}(C_{LM,1-1}^{L-1M'}\delta_{M'M-1}+C_{LM,11}^{L-1M'}\delta_{M'M+1})\}.\nonumber \\
\end{eqnarray}
Here, 
\begin{eqnarray}
&&S_{\mu} = -\sum_{i,j}u_i\{B^{-1}A'Z_{\mu}A'B^{-1}\}_{ij}u_j\\
&&S'_{\mu} = -\sum_{i,j}u'_i\{B^{-1}AZ_{\mu}AB^{-1}\}_{ij}u'_j\\
&&T_{\mu} = \sum_{i,j}u_i\{B^{-1}A'Z_{\mu}AB^{-1}\}_{ij}u'_j,
\end{eqnarray}
where, {\it e.g.} for $\mu = xy$,
\begin{eqnarray}
Z_{xy,jk} = -N\sum_{i=1}^{N_h} U_{ji}U_{ki}\delta_{\Gamma_{yz}\Gamma'_{i}}\delta_{\Gamma'_{i}\Gamma'_{zx}}\prod_{j\neq i}^{N_h}\delta_{\Gamma'_j\Gamma_j}
\end{eqnarray}
and $c_{11} = c_{10}/2$. Finally, we show the matrix elements of the Coulomb potential:
\begin{eqnarray}
&&\langle L'M',\bm{v}', A'|\frac{1}{r_{ij}}|LM,\bm{v}, A\rangle\nonumber \\
&&= \langle L'M',\bm{v}', A'|LM,\bm{v}, A\rangle\nonumber \\
&&\times\sum_{n=0}^{L}\frac{L!}{(L-n)!}\left( \frac{\sigma\sigma'}{c\rho}\right)^n\sqrt{\frac{2c}{\pi}}\frac{(-1)^n}{(2n+1)n!},
\end{eqnarray}
where
%\begin{equation}
%c^{-1} = \sum_{k,l}w^{(ij)}_k \{B^{-1}\}_{kl}w^{(ij)}_l, 
%\end{equation}
%\begin{equation}
%\sigma = c\sum_{k,l}w^{(ij)}_k \{B^{-1}\}_{kl}u_l,
%\end{equation}
%\begin{equation}
%\sigma' = c\sum_{k,l}w^{(ij)}_k \{B^{-1}\}_{kl}u'_l,
%\end{equation}
%\begin{equation}
%w^{(ij)}_k = \{U^{-1}\}_{ik}-\{U^{-1}\}_{jk}.
%\end{equation}
\begin{eqnarray}
&&c^{-1} = \sum_{k,l}w^{(ij)}_k \{B^{-1}\}_{kl}w^{(ij)}_l, \\
&&\sigma = c\sum_{k,l}w^{(ij)}_k \{B^{-1}\}_{kl}u_l,\\
&&\sigma' = c\sum_{k,l}w^{(ij)}_k \{B^{-1}\}_{kl}u'_l,\\
&&w^{(ij)}_k = \{U^{-1}\}_{ik}-\{U^{-1}\}_{jk}.
\end{eqnarray}

%\begin{acknowledgments}
%\end{acknowledgments}


\end{document}